\documentclass[]{aa}

\usepackage{txfonts}
\usepackage{float}
\usepackage{graphicx}
\usepackage{subfigure}
\usepackage{tabularx}

\begin{document}

\title{The XXL Survey VIII: MUSE characterisation of intracluster light in a z$\sim$0.53 
cluster of galaxies
~\thanks{Based on observations made with ESO Telescopes 
    at the La Silla and Paranal Observatories under programmes ID 191.A-0268
    and 60.A-9302. 
}
}

\author{
C.~Adami\inst{1} \and
E.~Pompei\inst{2} \and
T.~Sadibekova\inst{3} \and
N.~Clerc\inst{4} \and
A.~Iovino\inst{7} \and
S.L.~McGee\inst{9} \and
L.~Guennou\inst{6,1} \and
\\
M.~Birkinshaw\inst{5} \and
C.~Horellou\inst{13} \and
S.~Maurogordato\inst{8} \and
F.~Pacaud\inst{10} \and
M.~Pierre\inst{3} \and
B.~Poggianti\inst{11} \and
J.~Willis\inst{12}
}

\offprints{C. Adami \email{christophe.adami@lam.fr}}

\institute{
Aix Marseille Universit\'e, CNRS, LAM (Laboratoire d'Astrophysique de Marseille) 
UMR 7326, F-13388, Marseille, France 
\and
European Southern Observatory, Alonso de Cordova 3107, Vitacura, 19001 Casilla, Santiago 19, Chile
\and
Laboratoire AIM, CEA/DSM/IRFU/SAp, CEA Saclay, 91191, Gif-sur-Yvette, France 
\and
Max-Planck-Institute for Extraterrestrial Physics, Giessenbachstrasse 1, D-85748, Garching, Germany
\and
HH Wills Physics Lab., Univ. of Bristol, Tyndall Ave., Bristol BS8 1TL, U.K.
\and
Institut d'Astrophysique Spatiale (IAS), bat. 121, F-91405 Orsay Cedex, France
\and
INAF - Osservatorio Astronomico di Brera, via Brera, 28, I-20121 Milano, Italy
\and
Laboratoire Lagrange, UMR7293, Université de Nice Sophia Antipolis, CNRS, Observatoire de la Côte d'Azur, F-06300, Nice, France
\and
School of Physics and Astronomy, University of Birmingham, Edgbaston, Birmingham, B15 2TT, U.K.
\and
Argelander Institut f\"ur Astronomie, Universit\"at Bonn, Auf dem H\"ugel 71, D-53121 Bonn, Germany
\and
INAF-Astronomical Observatory of Padova, Italy
\and
Department of Physics and Astronomy, University of Victoria, 3800 Finnerty Road, Victoria, BC, Canada
\and
Dept. of Earth $\&$ Space Sciences, Chalmers University of Technology, Onsala Space Observatory, SE-439 92 Onsala, Sweden
}

\date{Accepted . Received ; Draft printed: \today}

\authorrunning{Adami et al.}

\titlerunning{MUSE characterisations of Intra Cluster Light in a z$\sim$0.53 cluster of galaxies}

\abstract 
{}
{Within a cluster, gravitational effects can lead to the removal of stars from
their parent galaxies and their subsequent dispersal into the intracluster medium. Gas
hydrodynamical effects can additionally strip gas and dust from galaxies; both gas and stars contribute to 
intracluster light (hereafter ICL). The properties of the ICL can therefore help constrain the physical processes at work in clusters 
by serving as a fossil record of the interaction history.}
{The present study is designed to characterise this ICL for the first time in 
a $\sim$10$^{14}$M$_\odot$ and z$\sim$0.53 cluster of galaxies  from  imaging and  spectroscopic points of view.
By applying a wavelet-based method to CFHT Megacam and WIRCAM images, we detect significant quantities of diffuse light and are able to 
constrain their spectral energy distributions. These sources were then spectroscopically characterised with 
ESO Multi Unit Spectroscopic Explorer (MUSE) spectroscopic data. MUSE data were also used to compute redshifts of 24 cluster galaxies and search for cluster substructures.}
{An atypically large amount of ICL, equivalent in i' to the emission from two brightest cluster galaxies, has been detected in this cluster. 
Part of the detected diffuse light has a very weak optical stellar component and apparently consists mainly of gas emission, while other 
diffuse light sources are clearly dominated by old stars. Furthermore, emission lines were 
detected in several places of diffuse light. Our spectral analysis shows that this emission likely originates from low-excitation parameter gas.
Globally, the stellar contribution to the ICL is about 2.3$\times$10$^9$ yrs old even though the ICL is not currently forming a large number of 
stars. On the other hand, the contribution of the gas emission to the ICL in the optical is much greater than the stellar contribution in some 
regions, but the gas density is likely too low to form stars. These observations favour  ram pressure stripping, turbulent viscous stripping, or 
supernovae winds as the origin of the large amount of intracluster light. \\
Since the cluster appears not to be in a major merging phase, we conclude that ram pressure stripping is the most 
plausible process that generates the observed ICL sources.}
{This is one of the first times that we are able to spectroscopically study diffuse light in such a distant and massive cluster, and it 
demonstrates the potential of MUSE observations for such studies. }


\maketitle

\section{Introduction}

The diffuse light within a galaxy cluster, often called intracluster light (ICL), is a fossil record of the formation, 
accretion, and interaction history of the cluster (e.g. Guennou et. al., 2012). It has been known for some time  that 
galaxies in clusters have different properties (morphology, star formation rate, colour, etc.) from several galaxies of
similar mass
in the field (Dressler et al. 1980, 1997; Weinmann et al. 2006). Physical processes could 
drive these differences, but luckily, as galaxies are accreted and orbit within a cluster, each physical process 
may leave characteristic low surface brightness signatures. Gravitational effects, including mergers and low-energy interactions 
(Mihos 2004, 2005), tidal stripping (Henriksen \& Byrd, 1996), and high-speed harassment (Moore et al. 1998), will strip 
or eject stellar material into the intracluster medium (ICM). This was recently statistically studied within the CLASH 
sample (Postman et al. 2012) by Burke et al. (2015), who concluded that the growth of stellar mass in the ICL is larger than can be 
provided by the brightest cluster galaxy (BCG hereafter) close 
companions, and that the majority of the ICL mass must come from galaxies which fall from outside of the core of the clusters.
This would be in agreement with De Maio et al. (2015) who used the same sample to conclude that the ICL is built up by the stripping of  
galaxies that are not too faint, and disfavoured significant contribution to the ICL by dwarf disruption or major mergers with the BCG. 
Alternatively, hydrodynamical effects such as ram pressure 
stripping of cold disk gas and dust (Gunn \& Gott 1972) can also pollute the ICM with gas that may become the material of future 
in situ star formation. We note, however, that studies such as Melnick et al. (2012) clearly do not favour a very intense 
in situ star formation. Excitation processes within this gas may lead to detectable emission lines within the ICL. 

Current observations of ICL have shown that there are large stellar streams that provide one viable input mechanism 
to the ICL (Rudick et al., 2009), but only a handful of such streams are known (e.g. Coma, Centaurus, Virgo, Perseus, Norma) and 
 the total amount of ICL is almost always equivalent at maximum to a few L$_\star$ galaxies in massive clusters (e.g. Guennou 
et al. 2012). From a general point of view, the ICL contribution to the cluster luminosity is less than a few dozen percentage points.
For example, using Type Ia supernovae, Sand et al.
(2011) estimated in a large cluster sample the intracluster (intergalactic) stellar mass to be of the order of 16$\%$ of the total
cluster stellar mass. 
Gonzalez et al. (2007) found a value of 19$\%$ when expressed with the similar conventions as 
in Sand et al. (2011). 

Most of the literature studies of the ICL are based on imaging surveys. There are, however, some noticeable exceptions,  
for example in Melnick et al. (2012). These authors used a z=0.29 cluster of galaxies (RXJ0054.0-2823) and stacked several spectra from
four different places in the cluster; however, this study was  forced to use the ICL and the BCGs together and to rely on 
the surface brightness profiles to delineate the boundaries between the two components. The ICL component was assumed to be
located at more than $\sim$50 kpc from the BCGs (but a significant contamination by a BCG is still possible). They found 
that the majority of the ICL stars in this cluster are old, but with high metallicities probably related to the fact that the 
considered cluster has three interacting giant elliptical galaxies.  Interestingly, they also  found a 408$\pm$20 km/s velocity 
dispersion for the ICL very close to the cluster velocity dispersion itself and significantly larger than the BCG velocity 
dispersion. This also puts  the BCG-independent nature of the ICL on  firmer ground. They finally detected extremely weak [OII] and [OIII] 
emission lines with intensity ratios that might be consistent with those of metal rich HII regions.

The present paper is based on the detection in the XXL Survey (Pierre et al. 2015, hereafter XXL paper I) of a cluster,  XLSSC 116, at 
z=0.534 with an exceptional amount of ICL  compared to other studied clusters in the literature; in this cluster this is equivalent to almost 
two brightest cluster galaxies (BCG hereafter) in the i' band, which is nearly ten times larger than normally observed. 
The existence of such a system is puzzling in the common scenarios of ICL formation (e.g. Guennou et al. 2012).  
While ICL is expected to be efficiently formed in clusters/groups, the amount observed in XLSSC 116 is exceptional 
and likely indicates recent strong dynamical interactions within the cluster. Such strong interactions may result from a 
high degree of substructure merging within a young cluster containing a low-temperature intracluster medium. Alternatively, 
such interactions may take place in a merging system with a currently shallow potential well, where the encounters 
are slower than normal owing to the lower characteristic velocity dispersion.

The present study is designed to characterise the ICL using both imaging and spectroscopy for the first time in 
a massive and distant cluster of galaxies. We will investigate the relative contribution of old stars, presumably 
stripped from the interacting galaxies, and the in situ intracluster star formation. While we 
expect that the first contribution will be dominant, the second seems to have been observed in Virgo (Verdugo et al. 2015).

We first describe the cluster from an X-ray point of view. In Section 3 we discuss the optical broadband data and the 
subsequent detection/characterisation of the diffuse light. Sections 4 and 5 describe the Multi Unit Spectroscopic Explorer (MUSE) measurements 
and the characterisation of the cluster ICL. Finally, we conclude in Section 6.  

We adopt the standard concordance cosmological model (H$_0$=72~km~s$^{-1}$~Mpc$^{-1}$, $\Omega _\Lambda = 0.74$,
$\Omega_M=0.26$).

\section{ XLSSC 116  from an X-ray point of view}

Our target is a cluster of galaxies detected in the XXL Survey (XXL paper I) as a significant extended 
X-ray source. This is not one of the most
significant clusters (only C2 class, see XXL paper I), and is not part of the 100
brightest XXL clusters (see Pacaud et al. 2015, hereafter XXL paper II \footnote{
XXL 100 brightest cluster data are available in computer readable form via the XXL Master Catalogue
browser http://cosmosdb.iasf-milano.inaf.it/XXL}). However, it visually exhibits a very large
amount of diffuse light (see below) and so has probably experienced an unusual evolution. 
This source is probably affected by non-thermal emission because the galaxy dominating the cluster 
is detected as a radio source (potentially coinciding with NVSS J021039-055637, later 
identified with FIRST data at 02h10m39.7sec -05deg56$\arcmin$36.9$\arcsec$ with a possible 
extension of $\sim$1.5$\arcsec$), and it was confirmed as a
cluster at z$\sim$0.53 (with the redshift of the BCG) in the course of the general XXL 
follow-up (WHT observations; see Koulouridis et al. 2015, hereafter XXL paper XII, for details on observations and data reduction).
It was previously detected as a candidate cluster in the optical with a photometric redshift of 0.47 by
Durret et al. (2011: W1-0957).

\subsection{X-ray morphology of the XLSSC 116 cluster}

\begin{figure}[h]
\includegraphics[width=\linewidth]{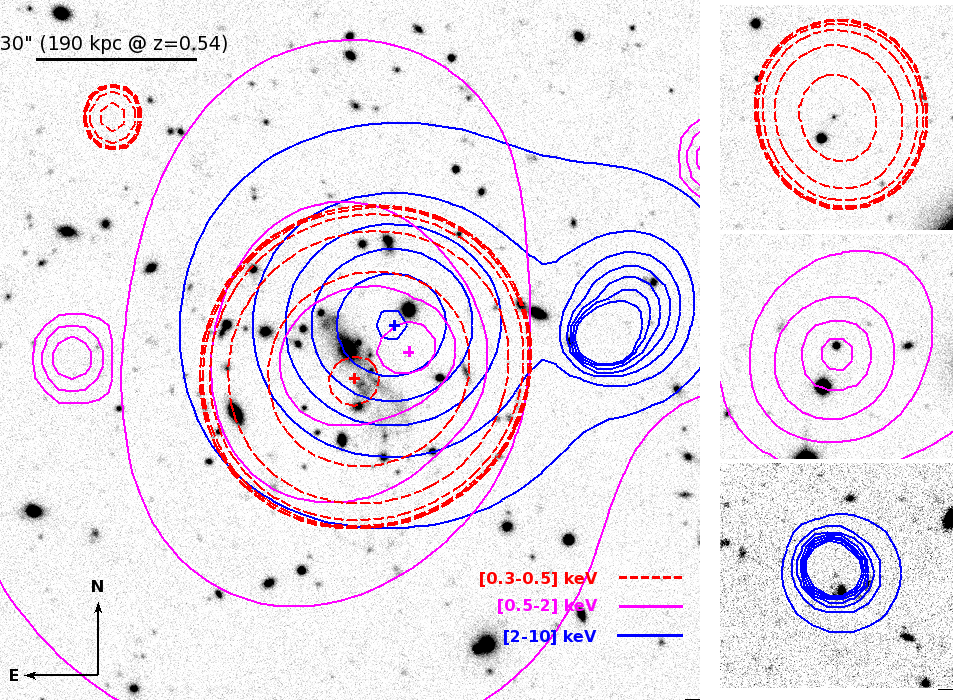}
\includegraphics[width=\linewidth]{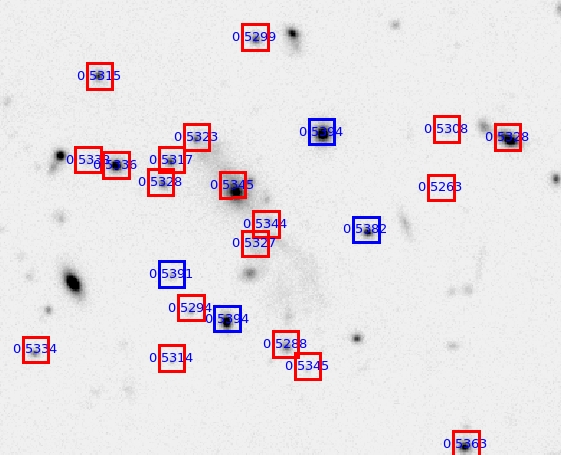}
\caption{\label{fig:rgb_xray} Upper figures:  X-ray morphology of XLSSC 116 as a function of energy band. The image is 
CFHT-LS $i'$ with X-ray contours overlaid. Crosses indicate the position of the surface brightness peak in 
each selected sub-band. The right panels show 
three point-sources in the vicinity of XLSSC 116 with spatial scales and contour levels matching those of the 
main image. They illustrate the size of the XMM-Newton PSF at the location of the cluster. We note the presence of a 
point-source in the [2-10]~keV band west of the cluster centre. This source is masked out in the X-ray analyses.
Lower figure: zoom on the same $i'$ image with Serna-Gerbal substructure 1 shown as red squares and substructure 2 
shown as blue squares (see below). Galaxy redshifts are also shown.}
\end{figure}

Figure~\ref{fig:rgb_xray} shows surface brightness contours of the galaxy cluster in three X-ray 
bands: [0.3-0.5], [0.5-2] (usual soft energy band encompassing most of the Bremstrahlung 
emission) and [2-10]~keV. Despite the limited number of X-ray counts in each band and 
the size of the XMM point-spread function (PSF), surface brightness contours may reveal a bimodality
of the extended emission. Lower-temperature gas (more prominent in [0.3-0.5]~keV) is shifted 
with respect to the hotter phases. The [0.5-2]~keV 
contours are elongated and show an 11~$\arcsec$ offset between the surface brightness peak and the 
BCG ($\sim$70 kpc at the clusters redshift). 
In order to put error bars on the separation, we randomly eliminate one thousand times 50$\%$ of the counts 
in the 0.5-2 keV bandpass and recalculate the centroid each time. The distribution of separations in these trials 
gives a measure of the uncertainty. Assuming Poisson statistics, the 1$\sigma$ uncertainty using all counts is 
$\sim$5$\arcsec$. This 11$\pm$5~$\arcsec$ shift is relatively important (e.g. Adami
$\&$ Ulmer 2000), potentially indicating that the cluster has a complex formation history. The other 
considered X-ray bands are also shifted with respect to the BCG position, but the shifts are smaller and not significant
given the data available. Unless the X-ray emission is contaminated by a faint, unresolved, 
point-source, this peculiar morphology may support the picture of different gas phases in the process of merging. We 
 note, however,  that the typical intrinsic XMM PSF width where the cluster 
is located is of the order of 8~$\arcsec$, resulting in a 9.4$\arcsec$ total uncertainty at this place. Higher resolution X-ray data 
(Chandra) would therefore be necessary to more efficiently discriminate between the different regions presently considered.

\subsection{ X-ray spectral analysis of the XLSSC 116 cluster}

        The X-ray spectrum is extracted in a region of radius 90$\arcsec$ around the BCG position. 
This corresponds to slightly larger than $r_{500}$ (see below), that includes the densest parts of the cluster.
It contains 806 net counts (thermal + non-thermal emission) in [0.3-10]~keV. We used {\sc XSpec} (Arnaud 1996) to 
compute best-fit  values and confidence intervals.

        We first fit the spectrum with a single-temperature APEC plasma model, fixing the metallicity abundance 
to 0.3 solar (Grevesse \& Sauval, 1998), and we obtain a best-fit temperature of $2.0_{-0.5}^{+1.3}$~keV. Since 
$r_{500} \sim 520 {\rm kpc} \equiv 83$~$\arcsec$ at the cluster redshift (from Sun et al. (2009) $T-{r_{500}}$ relation, and 
assuming $T=2$~keV), this temperature measurement roughly corresponds to $T_{500}$.
        
        If we perform a similar fit with the metallicity parameter (abundance) left free, we find a best-fit at 
$T=2.1_{-0.7}^{+1.3}$~keV and $Ab=0.3$ (abundance in solar units), but we cannot assign error bars to the metallicity parameter given the  
uncertainties in the fit that are too large.
        
\begin{figure}[!ht]
\begin{center}
\includegraphics[angle=0,width=3.5in,clip=true]{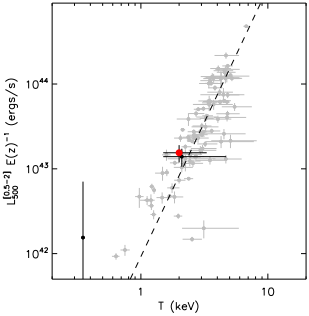}
\caption{Position of XLSSC 116 in the soft-band luminosity versus temperature plane (red dot with error bars). Grey points 
are measurements from the XXL 100 brightest cluster sample (Giles et al. 2015, hereafter XXL paper III). The straight line represent the scaling 
relation from XXL paper III fitted to the XXL 100 sample (forced to self-similar evolution). Black dots with error bars correspond 
to the two-component APEC fit to the cluster spectrum, possibly indicating the presence of two structures. The low temperature point 
has no temperature error-bar because we cannot constrain it owing to the small number of photons.}
\label{fig:plot_lt}
  \end{center}
\end{figure}

        A double-component APEC model fit with different temperatures 
and normalisations, both at fixed $Ab=0.3$ solar, gives a valley of minima for 
temperatures $T_1 \sim 2$~keV (resp. $T_2 \sim 2$~keV), corresponding to the previous case where the normalisation 
of component 2 (resp. component 1) is set to zero.
        More interestingly, we find a best-fit solution at $T_1=0.3$ and $T_2=2.1$~keV (the two black dots with error bars
in figure~\ref{fig:plot_lt}). Although the significance is low, 
this finding again supports the case for a bimodal gas distribution, with one cooler substructure merging into a hotter, 
2~keV cluster. According to these spectral models, the cooler component accounts for 10\% of the total system luminosity 
within a 90~$\arcsec$ (560~kpc) projected radius.
        

We measure a [0.5-2]~keV count-rate of $0.0198$ cts.s$^{-1}$  ($\pm 23\%$) in this radius, which translates into a 
galactic absorbed flux $f_{[0.5-2]}=1.8 \times 10^{-14}$~ergs/s/cm$^2$ ($\pm 23\%$). 
This value corresponds to a rest-frame [0.5-2]~keV luminosity $L_{500}^{[0.5-2]}=2.0 \times 10^{43}$ ergs/s and a 
bolometric luminosity $L_{500}^{bol}=4.7 \times 10^{43}$ ergs/s.
The position of the cluster in the luminosity-temperature plane is fully compatible with recent findings from the 
literature (Fig.~\ref{fig:plot_lt}). 

\section{CFHTLS and WIRCAM broadband data: diffuse light detection and spectral energy distribution fitting}

\subsection{Optical imaging data}

The cluster is located in the CFHTLS W1 field, providing u*, g', r', i', and z' data. We 
refer the reader to Coupon et al. (2009) for details. The limiting magnitude
of these images is u*$_{AB}$=26.3, g'$_{AB}$=26.0, r'$_{AB}$=25.6, i'$_{AB}$=25.4, z'$_{AB}$=25.0, and the
the pixel scale is 0.187$\arcsec$. In addition, we collected our own CFHT/WIRCAM K$_s$ data (subject
of a future paper). Briefly, they were obtained as 2$\times$525 seconds pointings with each 
pointing consisting of 21 dithers with 25 seconds of exposure per dither, for a total exposure time of 1050s.
Data were reduced by the standard TERAPIX pipeline and presented as 1 deg$^2$ tiles projected astrometrically 
onto CFHTLS optical images (same pixel scale). The depth is close to K$_s$ $_{AB}$=22 (5$\sigma$ level) in a Kron-type aperture.

\begin{figure}[!ht]
  \begin{center}
    \includegraphics[angle=270,width=3.7in,clip=true]{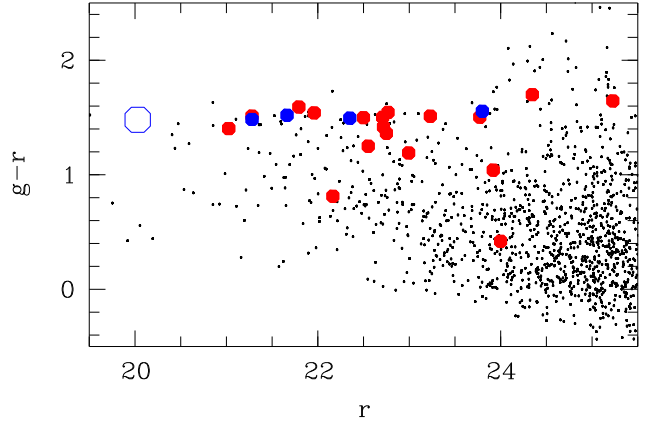}
    \caption{g'-r' versus r' for the CFHTLS objects in a 5$\arcmin$ square (allowing  a $\sim$1 Mpc radius at the cluster redshift) around the 
XLSSC 116 cluster centre. Points are all the detected objects;
filled circles are the galaxies with a MUSE measured spectroscopic redshift (red: SG substructure1, blue: SG substructure2, see Section 4.2). The 
large blue octagon is the sum of all the diffuse light components in the cluster.}
  \label{fig:n0308CMR}
  \end{center}
\end{figure}

Fig.~\ref{fig:n0308CMR} shows the galaxy colour magnitude relation in a 5$\arcmin$ square around the XLSSC 116 cluster centre.
We clearly see a narrow red sequence. Only five galaxies that are spectroscopically classified as cluster members (see below) are bluer 
than this red sequence. Of these, two are close to the BCG, well within the diffuse light region. 
The BCG is the brightest cluster member galaxy, but it is very different from a typical cD galaxy, as we 
 show later using the MUSE spectra.

\subsection{Diffuse light detection}

To detect the diffuse light, we applied the same tool as in Adami et al. (2013) and Guennou et al. (2012). This is a
wavelet-based method (the OV\_WAV package, see e.g. Pereira 2003; Da Rocha \& Mendes de Oliveira 
2005). OV\_WAV is a multiscale vision model (e.g. Ru\'e \& Bijaoui 1997). We briefly outline the main steps of the method. After applying 
a wavelet 
transform to an observation, the tool identifies the statistically significant pixels in the wavelet 
transform space (at the 5$\sigma$ level in our case). To define the objects, it groups pixels in 
connected fields for each scale. After the construction of an inter-scale connectivity tree, fields with connected pixels across three 
or more scales are identified and associated with the objects. 

We detected small-scale objects (typically the galaxies) in the sky image to produce our object image. We 
considered characteristic scales between 1 and 512 pixels in wavelet space (the 1024 pixel scale as in
Guennou et al. 2012 did not provide any additional detection). The object image was 
then subtracted from the sky image to produce the residual image. This residual image includes 
both hidden features that are typically too faint to satisfy the wavelet first-pass thresholding 
conditions and the diffuse light patches, which are too faint to be detected as objects. We did not have to
use a second iteration of this process as in Guennou et al. (2012) as the data were of sufficient quality to detect all
objects of interest in the first pass.

In a second step we searched for what we call the significant ICL sources, i.e. extended low surface-brightness 
features in the residual image. These features were detected in this image by considering the pixels where 
the signal is larger than 2.5$\sigma$ with respect to 20 empty areas of the residual image. These sources were 
visually inspected to remove obvious numerical residuals of bright saturated Galactic stars or defects due to 
image cosmetics. These numerical residuals are described in Adami et al. (2013); briefly,
they can be modelled by Sinc functions, which are negligible in intensity in the present case because the considered 
point sources are faint.

\begin{figure*}[!ht]
  \begin{center}
    \includegraphics[angle=0,width=5.5in,clip=false]{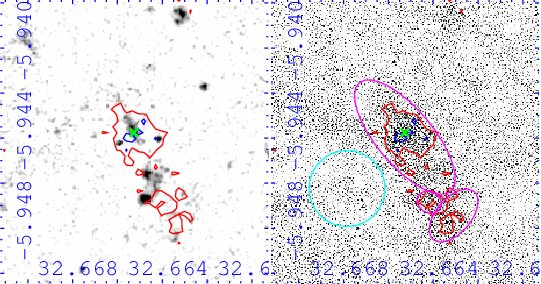}
    \includegraphics[angle=0,width=5.5in,clip=false]{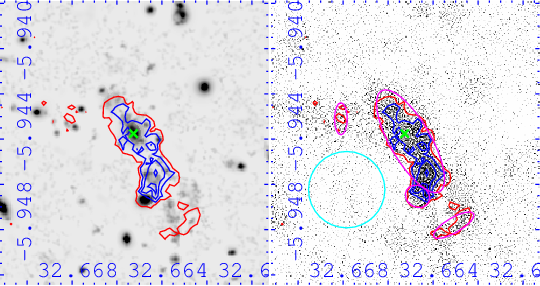}
    \includegraphics[angle=0,width=5.5in,clip=false]{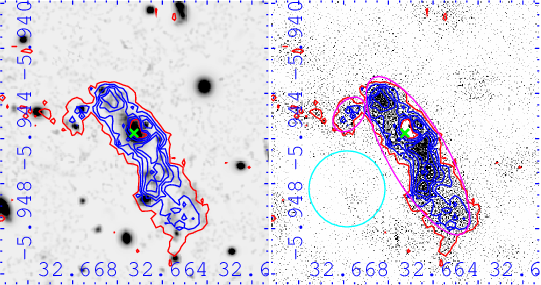}
    \caption{Left: original CFHTLS and WIRCAM images in the considered bands. Right: residual image from 
OV\_WAV. Red contour is the 2.5$\sigma$ level from the residual image. Blue contours start at the
3$\sigma$ level (from the residual image) and progress by steps of 1$\sigma$. These contours are computed with the 
residual images and reported on both residual and original images. The cyan circle (80kpc in diameter) shows one of the empty 
residual fields we considered. It also gives the physical size of the figure. Pink ellipses are the areas where we 
summed the flux of the diffuse light in the residual image. The green cross shows the BCG. From top 
to bottom: u*, g', r' bands.}
  \label{fig:images}
  \end{center}
\end{figure*}

\begin{figure*}[!ht]
  \begin{center}
    \includegraphics[angle=0,width=5.5in,clip=true]{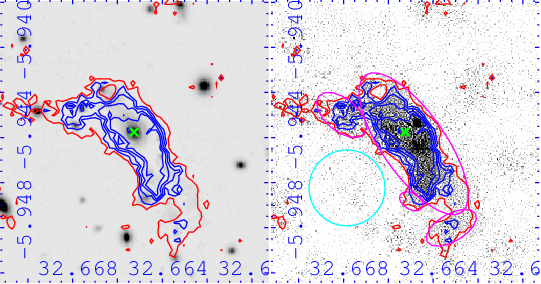}
    \includegraphics[angle=0,width=5.5in,clip=true]{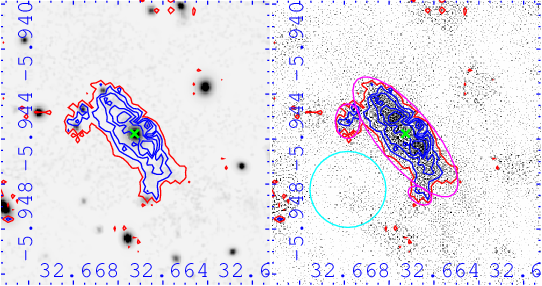}
    \includegraphics[angle=0,width=5.5in,clip=true]{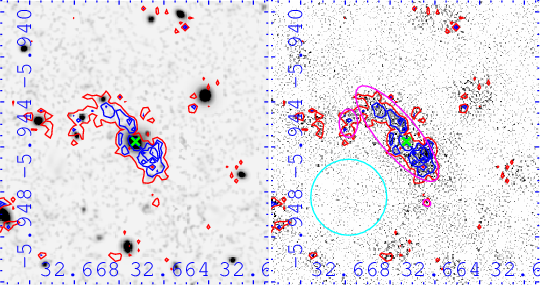}
    \caption{Same as fig.~\ref{fig:images} but for (from top to bottom): i', z', K$_s$ bands.}
  \label{fig:images2}
  \end{center}
\end{figure*}

This exercise was done independently with the u*, g', r', i', z', and K$_s$ images. We show in Figs.~\ref{fig:images}
and ~\ref{fig:images2} the results in the six available bands. Essentially, we detect a complex source of ICL  
in which several galaxies are embedded. This source (union of all ellipses in figs.~\ref{fig:images}
and ~\ref{fig:images2} and union of regions 1 and 2 in figs.~\ref{fig:figDL} and ~\ref{fig:figDL2}) is $\sim$60$\times$180 
kpc in size, which is very large 
compared to the typical ICL sources found by Guennou et al. (2012). Indeed, it is even larger than the plume detected 
in the Coma cluster by Gregg \& West (1998) or the diffuse light regions in Norma described in Fumagalli et al. (2014). 

We measured the source total magnitudes in the several available bands by integrating the flux of the residual images in the
pink ellipses in Figs.~\ref{fig:images} and ~\ref{fig:images2}. The size of these ellipses were adapted to the 
depth and quality of each band in order to best enclose the 2.5$\sigma$ detection area. We note that the fluxes
in the different bands were not measured  in exactly the same regions because the different considered bands do not have the same
depth or quality. For example, the u* and K$_s$ bands are much shallower than other optical bands, so considering exactly
the same areas would lead to including significant noise in our flux estimates. The magnitudes are compiled in Table 1, and clearly show 
an extraordinary amount of ICL. We compared these magnitudes to the magnitude of the BCG in each of five optical bands 
and found that the ICL has more than 2 times the flux of the BCG in g', r', i', or z'.

To the best of our knowledge, this is the first time such a system has been detected. Even the very high
concentration of ICL detected at z$\sim$2.1 in the CL J1449+0856 cluster (Adami et al. 2013) is far from being 
at the same level. Moreover, ICL in the XLSSC 116 cluster is clearly visible up to 130 kpc from the cluster centre,  
a feature that is also quite rare  in clusters.

We will now fit spectral energy distribution (SED hereafter) models on the diffuse light.

\begin{table*}[t!]
  \caption{Magnitudes (not corrected for galactic extinction,  which is lower than 0.1 magnitude in the less favourable case) of the 
detected ICL source and of the BCG from u* to K$_s$ wavelengths.}
\begin{tabular}{rrrrrrrr}
\hline
\hline
          &  u* AB  &  g' AB & r' AB & i' AB & z' AB & K$_s$ AB   \\
\hline
ICL & 23.58 $\pm$ 0.16 & 21.51 $\pm$ 0.14 & 20.03 $\pm$ 0.09 & 19.40 $\pm$ 0.02 & 19.27 $\pm$ 0.03 & 18.66 $\pm$ 0.05    \\
BCG & 23.339 $\pm$ 0.033 & 22.429 $\pm$ 0.011 & 21.025 $\pm$ 0.006 & 20.160 $\pm$ 0.004 & 19.823 $\pm$ 0.006 & 17.94 $\pm$ 0.05  \\
\hline
\end{tabular}
\label{tab:sample1}
\end{table*}

\subsection{SED fitting}

\begin{figure}[!ht]
  \begin{center}
    \includegraphics[angle=270,width=3.7in,clip=true]{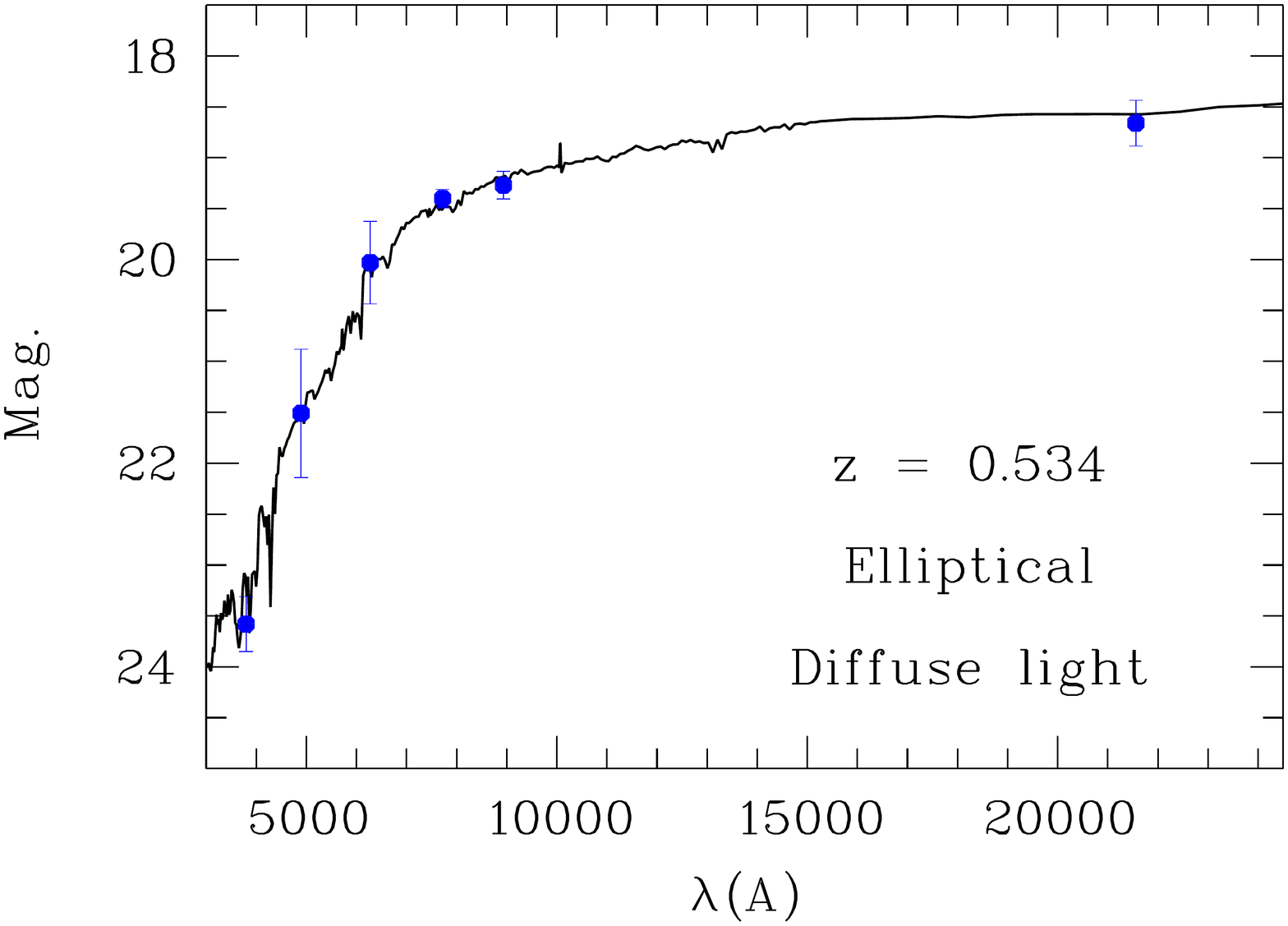}
    \includegraphics[angle=270,width=3.7in,clip=true]{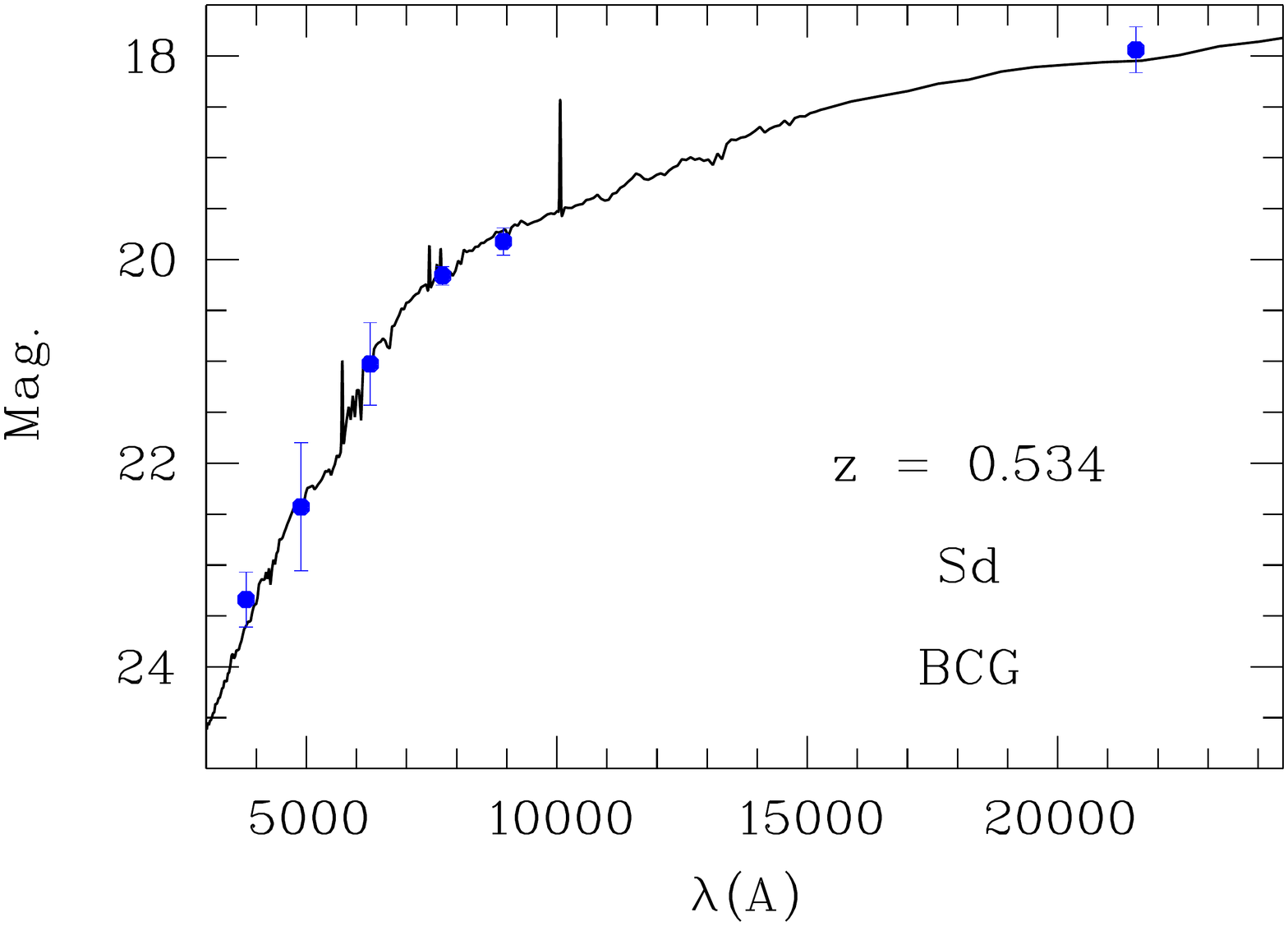}
    \includegraphics[angle=270,width=3.5in,clip=true]{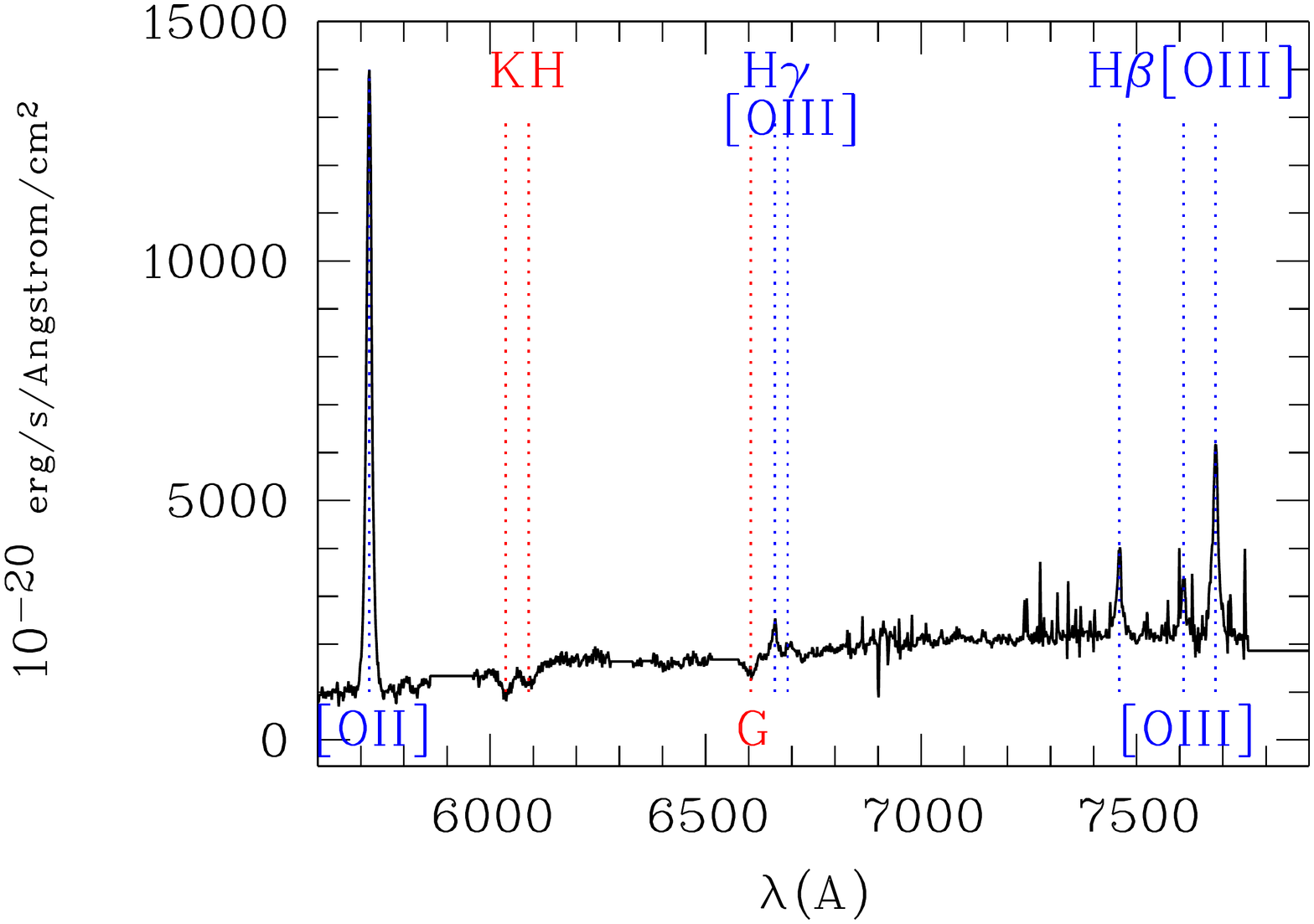}
    \caption{Spectral energy distribution fitted on the available u*, g', r', i', z', and K$_s$
magnitudes (3$\sigma$ error bars are also shown). Adopted redshift and best-fit model are given in each case. Upper figure: 
diffuse light, middle figure: BCG. The lower figure is the MUSE spectrum
of the BCG.}
  \label{fig:SEDKs}
  \end{center}
\end{figure}

The next obvious step is to fit a SED to the measured ICL magnitudes. We assume that the ICL source is related to the cluster and 
therefore has a redshift of 0.534 (confirmed in the following). We used the LePhare SEDfitting tool (Arnouts et al. 1999, Ilbert et 
al. 2006) and fixed the
redshift to the known value. We used the Cosmos survey templates (Ilbert et al. 2009) to estimate the closest galaxy type,
and the Bruzual $\&$ Charlot templates (2003) to estimate the stellar population age, stellar mass, and the star formation rate (SFR). We 
allowed discretised E(B-V) values between 0 and 1.0 with a 0.025 step. Uncertainty on E(B-V) was incorporated into the 
uncertainties given in Table 2. We also refer the reader to Ilbert et al. (2010: their Appendix D and section 4.1) for details on how the 
mass-to-light ratios were taken into account in the method. For reference, we also give  in Table 2 the resulting stellar-mass-to-light
ratios in the Ks band for the BCG and the ICL. We note here that the ICL stellar-mass-to-light ratio is only valid for the total amount
of ICL and not for its separate components (see below). We also note that we would need deeper Ks images in order to
have a more precise value of the Ks-band stellar-mass-to-light ratio for the ICL, given the very low surface brightness of the ICL features
(see e.g. fig.~\ref{fig:images2}).

In Fig.~\ref{fig:SEDKs} we show the best-fitting SED, and Table 2 gives the characteristics of the ICL. We see that the ICL
is well modelled by an early-type elliptical galaxy SED. We note that this does not conflict with having ICL emission lines 
in addition to this elliptical galaxy SED because wideband photometry is only weakly sensitive to narrowband emission lines.

The age of the ICL stellar population is 2.3 Gyr with an uncertainty allowing ages between 1 and 6 Gyr. The SFR
is poorly constrained and is likely of the order of 5 M$_\odot$ / yr, but could be as low as 0.07. 

\begin{table*}[t!]
  \caption{Characteristics of the ICL and of the BCG (type, age, stellar mass, SFR,  E(B-V) and stellar-mass-to-light
ratios in the Ks band). We give first the SED-estimated SFR, and then the spectroscopically estimated SFR.}
\begin{tabular}{rrrrrrr}
\hline
\hline
          &  Galaxy type &  Age       & log10(Stellar Mass) & log (SFR) & E(B-V) & log M/L K$_s$  \\
          &              &  10$^9$Gyr & M$_\odot$             & M$_\odot$ / yr &   & M$_\odot$ / L$_\odot$ \\
\hline
ICL & Ell & 2.3 [1.1; 5.9] & 10.7 [10.5; 10.9] & 0.7 [-1.2; 1.3] / 0.11 [0.04; 0.18] & 0.10$\pm$0.03 & 0.69 [0.52; 1.10]\\
BCG & Sd & 9.3 [7.9; 9.8] & 10.9 [10.4; 11.2] & 1.9 [1.5; 2.7] / 1.03 [0.96; 1.08] & 0.72$\pm$0.18 & 0.71 [0.25; 0.99]\\
\hline
\end{tabular}
\label{tab:sample2}
\end{table*}

A common criticism of ICL searches is that it is often difficult to discriminate between the BCG 
halo and the ICL itself. However, in the present case we note that given the ICL extension, there is no way to have such an extended 
BCG halo. Thus, it is likely that the diffuse light is only mildly contaminated by the BCG halo.

We have previously carried out several simulations (presented in Guennou et al. (2012) and Adami et al. (2013)) to be reasonably sure 
that our data 
are not heavily polluted by the underlying BCG. However, we perform an additional test in this paper by analysing the 
BCG in the same way as the ICL (see table 2 and Fig.~\ref{fig:SEDKs}). The flux from the BCG was considered
in its TERAPIX/CFHTLS ellipse: 3.5$\times$1.4$\arcsec$ ($\sim$ 23$\times$10 kpc) ellipse and with a trigonometric orientation of 311.7 degrees.
First, we note that the BCG is not passive (best type: Sd, high SFR). This is in good agreement with the galaxy  being   a radio source 
(probable) or  a UV source (possible): NVSS J021039-055637 and GALEXASC J021039.47-055645.5 are close to its 
position (respectively at $\sim$2$\arcsec$ and 9$\arcsec$). We also show  the MUSE spectrum (described below) of 
the BCG in Fig.~\ref{fig:SEDKs}. The spectrum clearly shows strong emission lines. This object is clearly
forming stars very actively. We do not have access to the H$\alpha$ line in the MUSE spectra, so we cannot directly measure the SFR from
this line. However, using the Moustakas et al. (2010) Sd galaxies (to be consistent with the best-fit type), we estimated an 
(H$\alpha$)/H$\beta$ ratio of 4.85$\pm$0.7. Measuring the flux under the H$\beta$ line in Fig.~\ref{fig:SEDKs}, translating it
into an H$\alpha$ flux, and using Kennicutt (1998), we find a spectroscopically estimated SFR of 10.6$\pm$1.5 M$_\odot$/yr. 
If we use the [OII] flux, the SFR estimate is 32.3$\pm$0.75 M$_\odot$ / yr.

\begin{figure}[!ht]
  \begin{center}
    \includegraphics[angle=270,width=3.7in,clip=true]{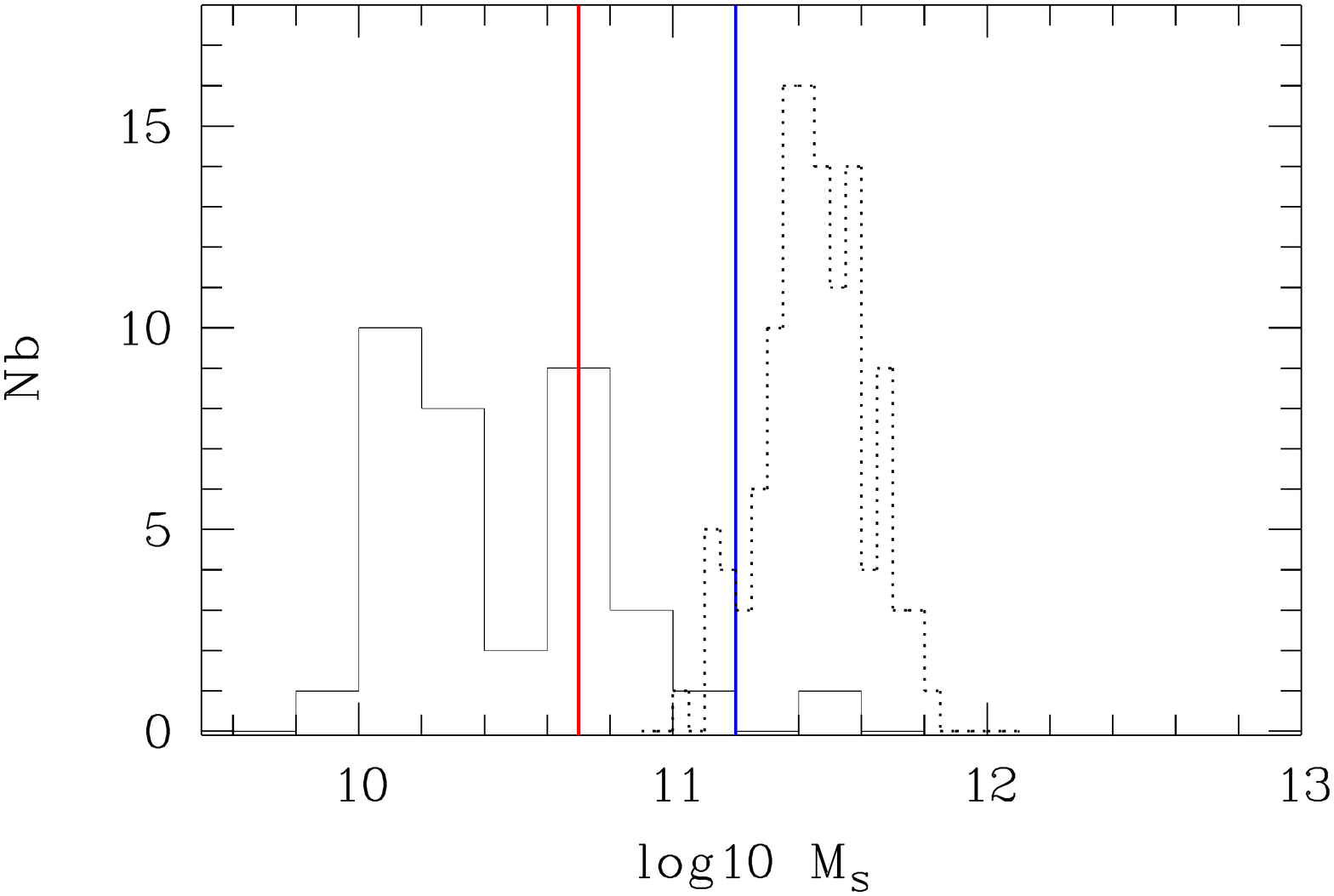}
    \includegraphics[angle=270,width=3.7in,clip=true]{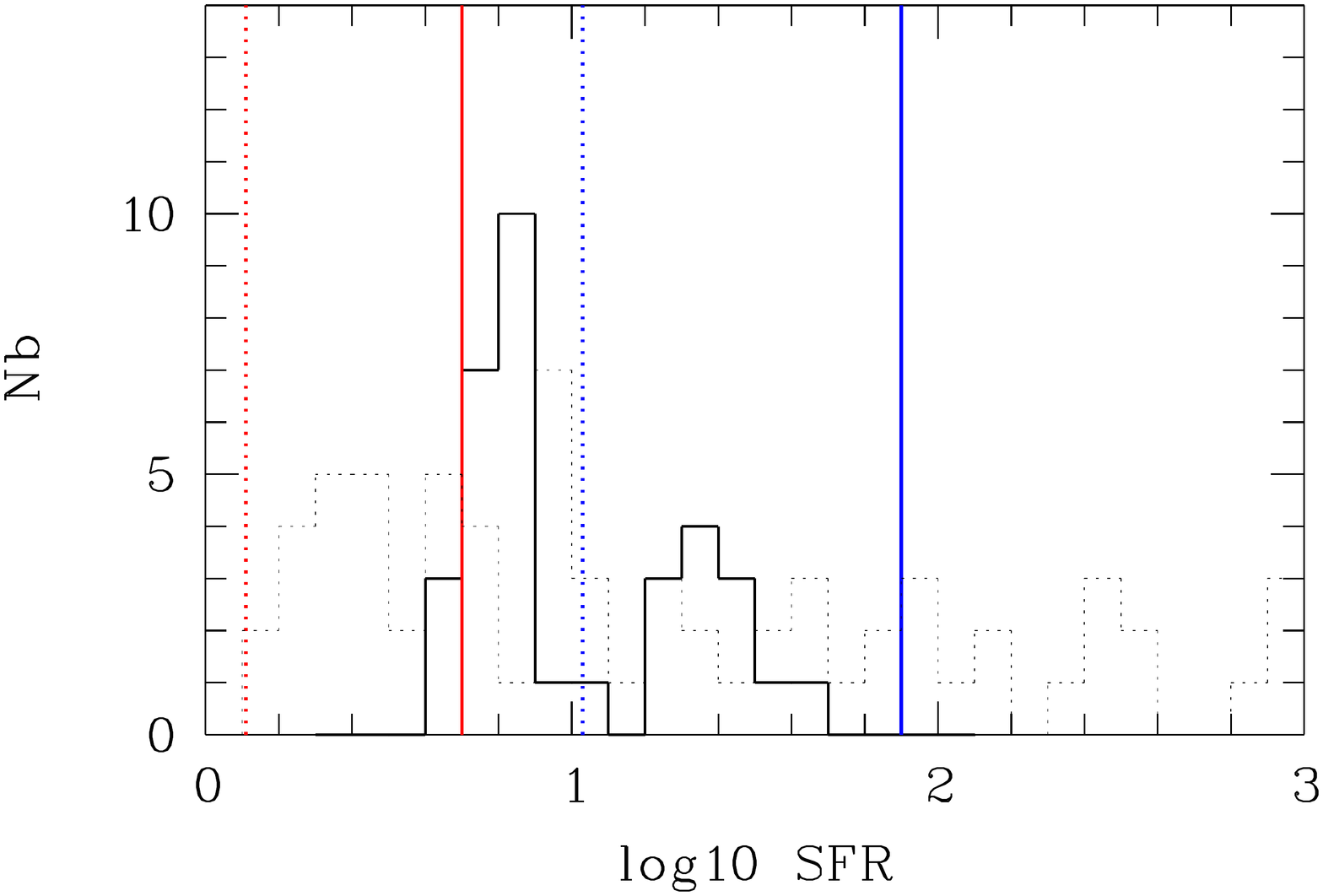}
    \caption{Comparison of our stellar mass and SFR estimates for the XLSSC 116 ICL source (red vertical line for SED estimation and red vertical 
dotted line for spectroscopic estimation) and BCG (blue
vertical line for SED estimation and blue vertical dotted line for spectroscopic estimation) with the Puech et al. (2010) emission line 
galaxies (continuous histograms) and the Liu et al. (2012) BCGs (dashed histograms). Upper figure: stellar mass, lower 
figure: SFR.}
  \label{fig:compar}
  \end{center}
\end{figure}

Given their different stellar population ages and SFRs, the ICL and BCG clearly have different stellar populations.
This is illustrated in Fig.~\ref{fig:compar}, where we have compared our SFR and stellar mass estimates with known literature values from Puech 
et al. (2010: emission line galaxies) and Liu et al. (2012: BCGs) at similar redshifts. The ICL source has a stellar mass typical of 
the emission line galaxy sample. The XLSSC 116 BCG has a stellar mass closer to the literature BCGs. Our ICL source seems 
to exhibit intermediate characteristics that are  between an old and passive elliptical galaxy and a younger and more active object. The 
XLSSC 116 BCG does not seem to be a normal cD galaxy with a relatively low stellar mass and a high star formation rate.

The estimates presented in this section are, however, only based on model fitting. We now discuss
the MUSE ICL spectra.

\section{MUSE data analysis}

\subsection{MUSE optical spectroscopic data and redshift measurement}

We were awarded four hours of MUSE Science Verification Time (\# 60.A-9302) in order 
to observe the cluster using integral-field spectroscopy. 
MUSE (see http://www.eso.org/sci/facilities/develop/instruments/muse.html)
is a second generation instrument for the Very Large Telescope (VLT) and is an integral-field spectrograph operating in the visible 
wavelength range. We obtained the data in the wide-field mode with adaptative-optic mode off, simultaneously covering a 1$\times$1 arcmin field 
from $\sim$480 to $\sim$930nm with a final spectral resolution of 1.25$\AA$ per pixel. The spatial resolution was of the order of 0.2$\arcsec$
 before convolution with the seeing, which varied between 0.85 and 1.2$\arcsec$ during the June run, and 
between 0.84 and 1.07$\arcsec$ during the August run.
The observations were centred on the XLSSC 116 BCG and executed in service mode 
during the nights of 25-26 June and 20-22 August 2014. We obtained four one-hour observing blocks,  each consisting of four exposures 
separated by 90$^{\circ}$ rotation, to average out 
the patterns of the slicers and channels seen by the detector.
The data were reduced following the recipes of the MUSE pipeline, version 0.18.2. 
The June data were taken at a temperature below 7$^{\circ}$, which caused problems with the 
correct identification of the 10th slice of IFU no. 6. To obviate the problem, we followed the
workaround recommended in the MUSE cookbook and used dedicated trace tables. 
The reduction steps for each individual exposure included bias, flat-field, wavelength and 
flux calibration, and correction for the telluric absorption lines using standard stars and 
geometric correction. Dedicated sky observations were taken in an area adjacent to the 
cluster and were reduced following the same recipe and were subtracted from the corresponding science data. The 16 final cubes 
were combined  using relative RA and DEC offsets, and keeping the whole wavelength range. The spectra were extracted by 
summing all the pixels in several elliptical regions of different 
extensions using ds9 (see Table 3 for a list of successful redshifts). 
Narrow- or broadband images can be obtained by collapsing the final cube in the 
selected wavelength range.

\begin{table}[t!]
  \caption{J2000 coordinates, redshifts, i' band magnitude, Serna-Gerbal substructure number of the CFHTLS
objects with a successful MUSE redshift measurement, and spectral quality flag (see text).}
\begin{tabular}{rrrrrr}
\hline
\hline
RA (2000)  &  DEC (2000)  &  z & i' & SG sub. & flag\\
\hline
32.6571 &  -5.9526 & 0.5363  & 21.06     & 1& 3\\
32.6556 &  -5.9416 & 0.5328  & 20.31     & 1& 4\\
32.6578 &  -5.9413 & 0.5308  & 23.43     & 1& 2\\
32.6580 &  -5.9434 & 0.5263  & 25.91     & 1& 2\\
32.6647 &  -5.9380 & 0.5299  & 21.69     & 1& 3\\
32.6703 &  -5.9394 & 0.5315  & 21.59     & 1& 3\\
32.6726 &  -5.9492 & 0.5334  & 22.31     & 1& 2\\
32.6623 &  -5.9414 & 0.5394  & 20.27     & 2& 4\\
32.6607 &  -5.9449 & 0.5382  & 21.40     & 2& 4\\
32.6628 &  -5.9498 & 0.5345  & 23.71     & 1& 2\\
32.6636 &  -5.9490 & 0.5288  & 21.89     & 1& 3\\
32.6677 &  -5.9495 & 0.5314  & 23.78     & 1& 2\\
32.6657 &  -5.9481 & 0.5394  & 20.73     & 2& 4\\
32.6670 &  -5.9477 & 0.5294  & 22.89     & 1& 2\\
32.6677 &  -5.9465 & 0.5391  & 23.01     & 2& 2\\
32.6647 &  -5.9454 & 0.5327  & 21.91     & 1& 2\\
32.6643 &  -5.9447 & 0.5344  & 22.31     & 1& 3\\
32.6655 &  -5.9433 & 0.5345  & 20.16     & 1& 3\\
32.6668 &  -5.9416 & 0.5323  & 21.89     & 1& 3\\
32.6681 &  -5.9432 & 0.5328  & 21.93     & 1& 2\\
32.6677 &  -5.9424 & 0.5317  & 21.74     & 1& 3\\
32.6697 &  -5.9426 & 0.5336  & 20.81     & 1& 4\\
32.6707 &  -5.9424 & 0.5338  & 23.15     & 1& 2\\
\hline
\end{tabular}
\label{tab:samplecfhtls}
\end{table}

The redshifts were obtained by using the EZ redshift measurement code (Garilli
et al. 2010) on the final 1D spectra, allowing an additional smoothing
of 3 pixels in order to find the redshift value more easily when needed. The redshift measurements were 
done in the same way as for the VIPERS survey (e.g. Guzzo et al. 2014) and a redshift measurement quality flag was
assigned between 1 and 4.  Flag 1 means that we have a 50$\%$  chance of having the correct
redshift estimate; flag 2, 75$\%$; flag 3, 95$\%$; and flag 4, more than
99$\%$.  We only considered  objects with flags 2, 3, and 4 as successful measurements. Statistically, this means that we may have 
three objects with an incorrect redshift in Table 3.

\begin{figure}[!ht]
  \begin{center}
    \includegraphics[angle=270,width=3.7in,clip=true]{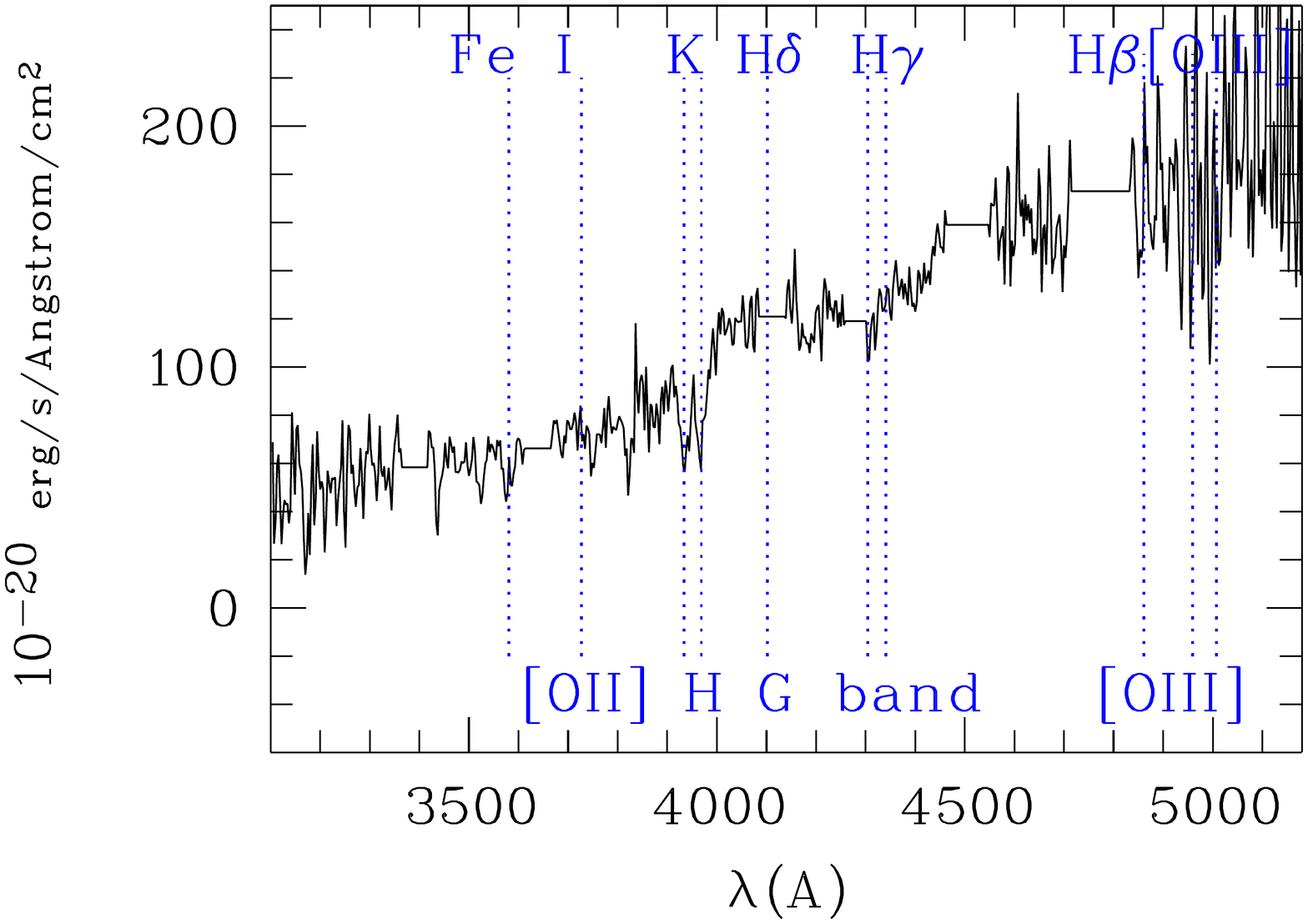}
    \includegraphics[angle=270,width=3.7in,clip=true]{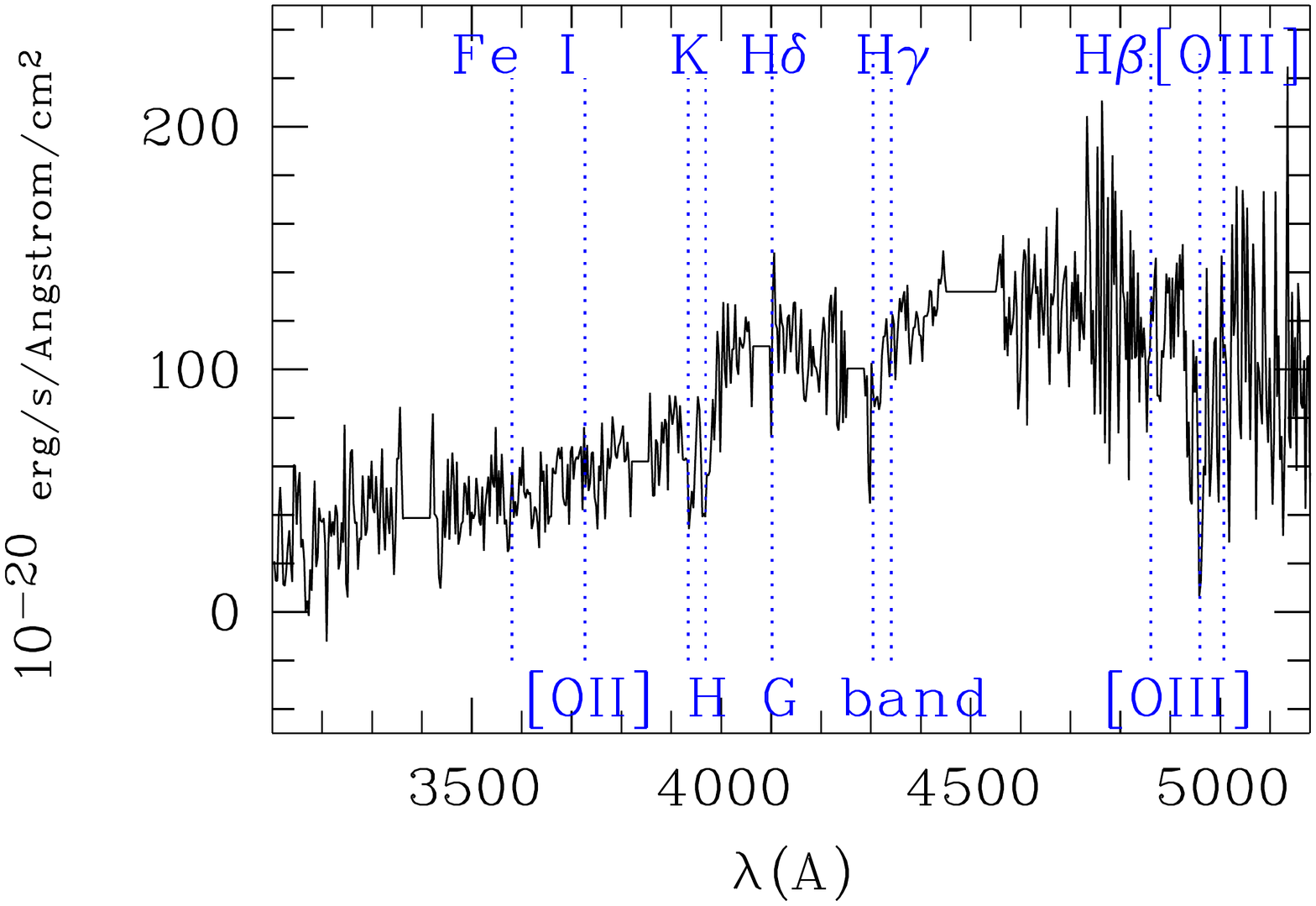}
    \caption{Mean (upper figure) and median (lower figure) rest frame spectrum of all the galaxies spectroscopically measured inside the cluster
(excluding the BCG). Strong sky lines have been masked. The position of the most common lines are shown as dashed lines. }
  \label{fig:synthe}
  \end{center}
\end{figure}

A detailed study of the individual galaxy spectra, and their comparison with other clusters detected in the XXL Survey, will be undertaken in a 
future paper following the completion of the survey spectroscopic follow-up. However,  in fig.~\ref{fig:synthe} we show the mean
and median spectra of all the galaxies (excluding the BCG) spectroscopically measured and found to be inside the cluster. Each spectrum was put 
in the rest frame based on the measured redshift and we computed the mean and median spectrum. This gives a mean and median view of the cluster 
galaxy population. This spectrum is nearly elliptical, with prominent absorption lines (e.g. H$\&$K), a strong Balmer break,
but without any significant emission lines.  We show later that the BCG has a very different spectrum. 
The completeness level, in terms of the percentage of galaxies with a successful redshift measurement, is 90$\%$ down to i'$\sim$20.5 and 
80$\%$ down to i'$\sim$22 in the MUSE field of view.

\subsection{Structure of the cluster}

Given the number of available spectroscopic redshifts, we can investigate the presence of possible substructures
in the cluster. To this end, we have applied the Serna-Gerbal (1996, hereafter SG) hierarchical method already extensively 
described in several papers (e.g. Guennou et al. 2014 and references therein). This method is quite powerful for showing 
evidence of substructure; the code also estimates the mass of the substructures. Masses are computed through a 
basic version of the virial theorem (neglecting electromagnetic 
fields and only using classical estimators for the galaxy velocity dispersion). These estimates suffer from relatively large 
uncertainties of the order of 4.6 10$^{13}$ M$\odot$. This value was statistically estimated from the data of Guennou et al. (2014)
by comparing the SG estimates and the masses deduced from the Giodini et al. (2009) cluster scaling relation. Guennou et al. (2014),
however, also showed that for a given parent-cluster the relative masses between substructures were still reliable. 

\begin{figure}[!ht]
  \begin{center}
    \includegraphics[angle=270,width=3.55in,clip=true]{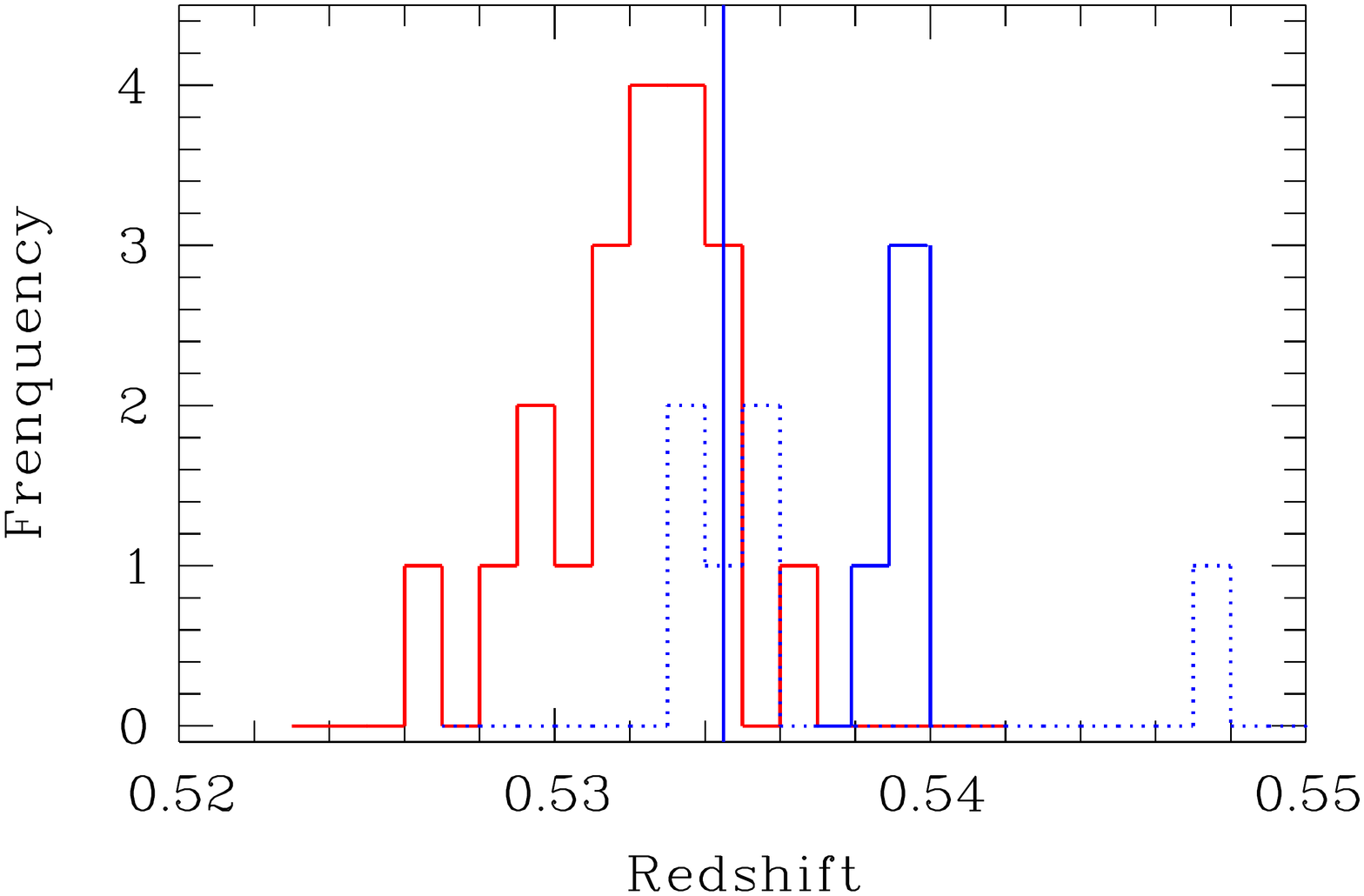}
    \caption{Redshift histogram of the MUSE CFHTLS galaxy redshifts inside the XLSSC 116 cluster (red solid histogram: SG substructure 1; 
blue solid histogram: SG substructure 2). We also show the non-CFHTLS-only emission line objects described in section 5.2 (blue dotted 
histogram). The vertical line is the redshift of the BCG.}
  \label{fig:galDL}
  \end{center}
\end{figure}

More precisely, the SG hierarchical method calculates the potential binding energy
between pairs of galaxies and detects substructures by taking positions, magnitudes, and redshifts into account. We required at least
four galaxies in a given substructure. The SG method detects two such substructures in the XLSSC 116 cluster (see table 4). The first substructure 
has 20 galaxies, an estimated mass of 3$\times$10$^{13}$ M$_\odot$, and a velocity dispersion of 570 km/s. The second substructure is
smaller: 4 galaxies, and the estimated mass and velocity dispersion are 4$\times$10$^{12}$ M$_\odot$ and 170 km/s. This second substructure probably 
corresponds to the low-temperature component detected in the X-ray data (the brightest galaxy of this substructure is very close 
to the X-ray peak of the low-temperature component; see figure~\ref{fig:rgb_xray}) and is at a greater distance ($\sim$2000 km/s) 
(see blue solid histogram in fig.~\ref{fig:galDL}).

 Galaxies that are members of the two 
detected substructures are also well mixed in the cluster red sequence (see 
fig.~\ref{fig:n0308CMR}). We  note that a velocity dispersion of 570 km/s in a relaxed cluster would suggest an X-ray 
temperature on the order of 2 keV (Rosati et al., 2002), in good agreement with our X-ray measurements.

\begin{table}[t!]
  \caption{Characteristics of SG detected substructures. Mass estimates are statistically affected by 4.6 10$^{13}$ M$\odot$ uncertainties.}
\begin{tabular}{rrrr}
\hline
\hline
SG number &  Number of galaxies & Vel. dispersion & Mass            \\
          &                     &  km/s               & 10$^{13}$M$_\odot$ \\
\hline
1 & 20 & 570 &   3 \\
2 &  4 & 170 & 0.4 \\
\hline
\end{tabular}
\label{tab:sample33}
\end{table}

\section{Nature of the cluster diffuse light with MUSE data}

The first question we want to answer concerning the diffuse light in clusters is whether there is any indication of line 
emitting gas. Does the old stellar population of diffuse light (as previously detected) also give rise to intergalactic 
gas ionisation? Or perhaps some collisional effects can ionise gas?

The first way to answer is to reconstruct narrowband images with spectral
data centred on major emission lines at the cluster redshift. Given the available MUSE data, we  built
[OII] (5699-5729 A), H$\beta$ (7432-7471 A), and [OIII] (7655-7696 A) images (lower and upper spectral bounds of these images were chosen 
approximately as the mean line wavelength $\pm$2 times the velocity dispersion of the main cluster structure detected below). 
This gives evidence of a new diffuse light region completely invisible in any of the 
broadband images directly south of the BCG and showing significant emission at several wavelengths. This emission line
diffuse light is denoted ELDL in the following.

\subsection{ ELDL source}

\begin{figure}[!ht]
  \begin{center}
    \includegraphics[angle=0,width=3.5in,clip=true]{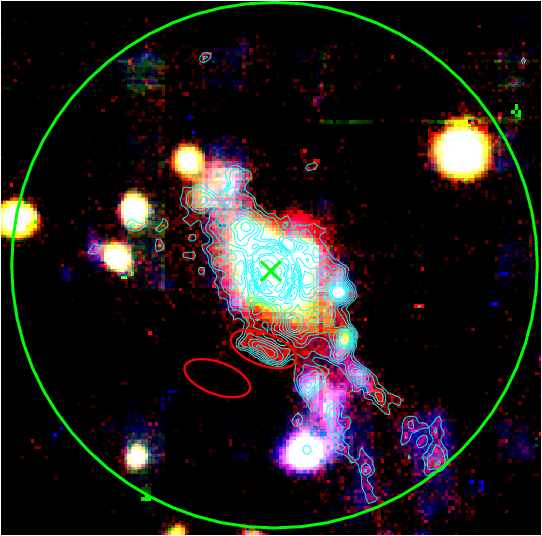}
    \includegraphics[angle=270,width=3.55in,clip=true]{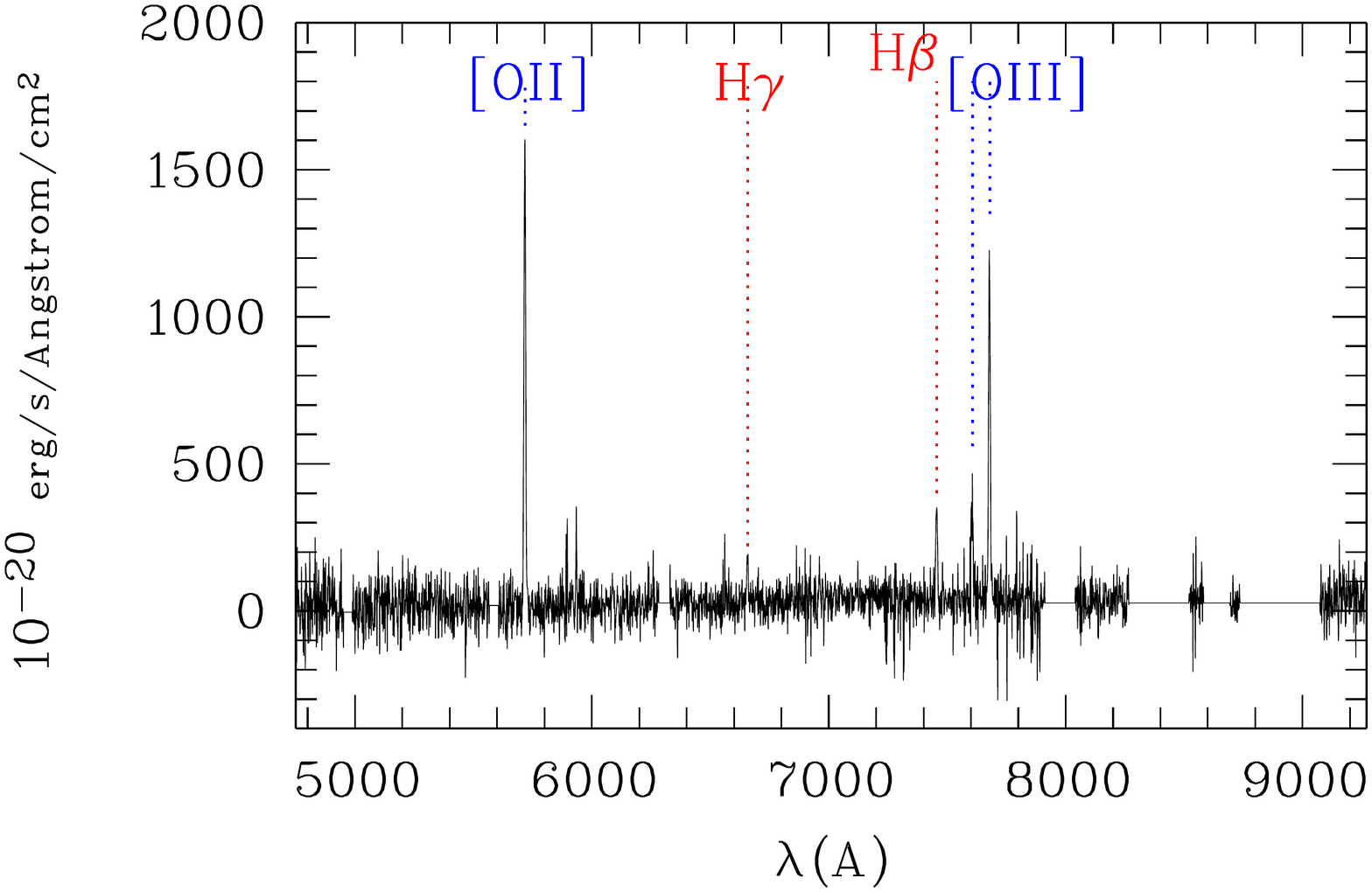}
    \caption{Upper figure: trichromic image of the ELDL area (blue is for V-band MUSE reconstructed
image, green for the I-band MUSE reconstructed image, and red is the narrowband [OII]+H$\beta$+[OIII] MUSE
reconstructed image). Upper red ellipse is the area where the 
spectrum of ELDL was extracted and lower ellipse is the region where the background was estimated. The large 
green circle is a 100kpc radius region and the green cross is the position of the BCG. [OII]+H$\beta$+[OIII] 
residual image from OV\_WAV is shown as cyan contours. These contours start at the 2.5$\sigma$ level and progress 
by steps of 0.5$\sigma$. Lower figure: calibrated spectrum of the ELDL.}
  \label{fig:emDL2}
  \end{center}
\end{figure}

The ELDL source (fig.~\ref{fig:emDL2}) mainly appears at two wavelengths (5713A and 7676A), exactly
centred on [OII] and [OIII] redshifted to the cluster redshift of 0.534. We also find a much weaker emission 
centred on H$\beta$. This is clearly a very low surface brightness object in 
terms of its continuum emission, which made it undetectable both in the CFHTLS and WIRCAM data (detection much lower 
than the 2$\sigma$ level in these data). We also estimated the significance of the ELDL source detection in the sum
of the MUSE [OII] (5699-5729 A), H$\beta$ (7432-7471 A), and [OIII] (7655-7696 A) images (see  fig.~\ref{fig:emDL2}).
The ELDL source is detected at the 5$\sigma$ level in these data, and most of the diffuse light sources previously described 
are detected at more than the 3$\sigma$ level. It is not the
first time emission lines have been detected in ICL sources (see e.g. Melnick et al. 2012); however, to the best of knowledge,
all other detections were several orders of magnitudes weaker in intensity for the oxygen lines.

We extracted a spectrum of this source inside the red ellipse in fig.~\ref{fig:emDL2}; the background was estimated and subtracted 
in a nearby region outside  the diffuse light regions
(second red ellipse). The background subtracted spectrum is shown in fig.~\ref{fig:emDL2} and 
continuum subtracted fluxes of the visible lines are given in Table 5.

\begin{table}[t!]
  \caption{ELDL continuum subtracted line [OII], H$\gamma$, H$\beta$, and [OIII] fluxes in units of 10$^{-20}$ erg/s/A/cm$^2$.}
\begin{tabular}{rrrrr}
\hline
\hline
 [OII] & H$\gamma$ & H$\beta$ & [OIII]a & [OIII]b   \\
 3727.3 A & 4340.5 A & 4861.3 A & 4958.9 A & 5006.8 A  \\
\hline
14420 & 12 & 2938 & 3526 & 10360  \\
\hline
\end{tabular}
\label{tab:ELDLflux}
\end{table}

These oxygen emission lines necessarily imply the presence of an ionised compact source of gas of $\sim$13$\times$6 kpc 
(the physical size of the red ellipses in fig.~\ref{fig:emDL2}). Other similar sources may be detected all around the 
BCG, but the one we  consider below is the only one to be clearly separated on the sky from the 
BCG halo. Using the MUSE data to reconstruct an I-band image (where the considered object is not
detectable), we estimated an upper value of the magnitude of this source of I$\geq$26.5, so M$_I$ $\geq$ -16 at z=0.534, roughly 
corresponding to a restframe B magnitude. Adopting B$_{\odot}$=5.48, this represents a maximum of 4 $\times$ 10$^8$ stars of solar type 
(and  probably much lower) leading
to a maximum stellar density of 0.0002 pc$^{-3}$. This is $\sim$500/1000 times lower than in the solar vicinity. This source is 
therefore extremely poor in terms of stars. 

\begin{figure}[!ht]
  \begin{center}
    \includegraphics[angle=270,width=3.55in,clip=true]{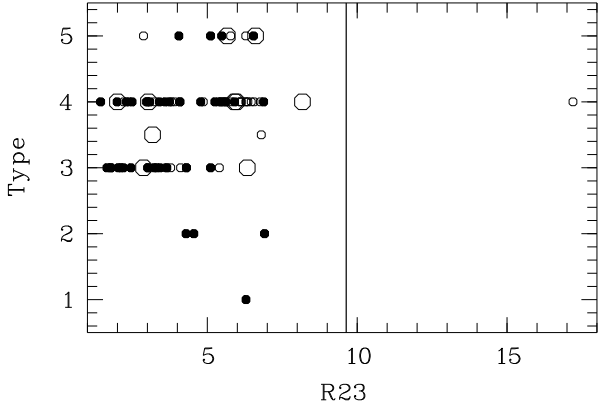}
    \caption{R23 versus galaxy morphological type in the Moustakas et al. (2010: filled circles) and Boselli 
et al. (2013: empty circles) samples. Large empty circles are active galaxies. The vertical line shows our ELDL source.}
  \label{fig:compargal}
  \end{center}
\end{figure}

At the same time, the [OII] and [OIII] emission of the considered source is comparable to the emission of the detected galaxies
around the source position, while the H$\beta$ flux is quite low. To compare the fluxes in Table 5 with known regular galaxies,
we selected the galaxies with a known precise morphological type and with measured [OII], H$\beta$, [OIII]a, and [OIII]b fluxes
in the samples of Moustakas et al. (2010) and Boselli et al. (2013). We computed the R23 parameter from Moustakas et al. (2010),
which is basically the ratio between the oxygen ([OII] and [OIII]) and the H$\beta$ emissions.  In fig.~\ref{fig:compargal}
we show that our ELDL source  clearly has a larger R23 than nearly all the galaxies in the two previous papers, whatever their morphological
type, origin (Moustakas et al. (2010) or Boselli et al. (2013)), or star-forming  activity. Oey $\&$ Shields (2000) show that only a 
few HII regions can show such a high R23 (see their Figure 7). The only models shown in their paper able to reproduce
such high values require very specific conditions for the gas: $\sim$0.3-0.5 solar metalicity and a low gas excitation parameter U 
of $\sim$0.01. Any of their models with metallicity larger than 1 predict R23 values at least 
five times smaller than what we observe in the ELDL source. 
We also compared our results with the work of Croxall et al. (2009) on HII regions in dwarf galaxies of the M81 group. From
their figure 2, only HII regions detected in tidal dwarf galaxies can reproduce our R23 value. This does not mean that the ELDL
source is a tidal dwarf galaxy as it is much larger ($\sim$13$\times$6 kpc ) than the typical size of tidal dwarf
galaxies. However, the gas included in the ELDL source probably has similar characteristics to those present in these dwarf galaxies. 

If we use the ELDL H$\beta$ flux in the same way as we did for the BCG to estimate a spectroscopic SFR, we have a low 
value of 1.3$\pm$0.2 M$_\odot$/yr. If we use the [OII] flux, the SFR estimate is 2.2 M$_\odot$ / yr.  This confirms that the 
ELDL source is not forming a large number of stars (or that it is very dusty).

Therefore, the ELDL is a galaxy-size, star-poor, and oxygen emitting source with a  of half solar metallicity and a low 
excitation parameter. This may be consistent with an extremely low surface brightness galaxy similar to the local group
dwarves of Gnedin (2000).


\begin{figure}[!ht]
  \begin{center}
    \includegraphics[angle=0,width=3.55in,clip=true]{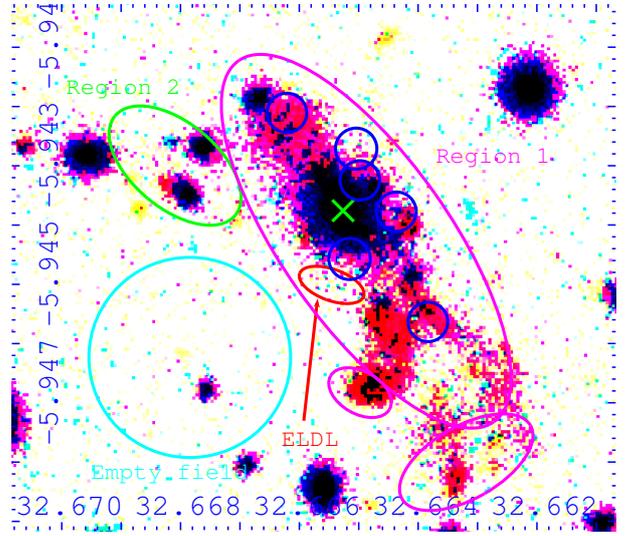}
    \caption{CFHTLS and WIRCAM g', i', and K$_s$ trichromic image of the cluster centre. Cyan circle: region where the 
background spectrum was extracted. It also gives the physical size of the figure. Diameter of this circle:
80kpc. Region 1: union of the three magenta ellipses. Region 2: green ellipse. Small blue circles: six 
detected non-CFHTLS emission line objects. Green cross: BCG position. Red ellipse: ELDL source position. }
  \label{fig:figDL}
  \end{center}
\end{figure}

\begin{figure}[!ht]
  \begin{center}
    \includegraphics[angle=270,width=3.55in,clip=true]{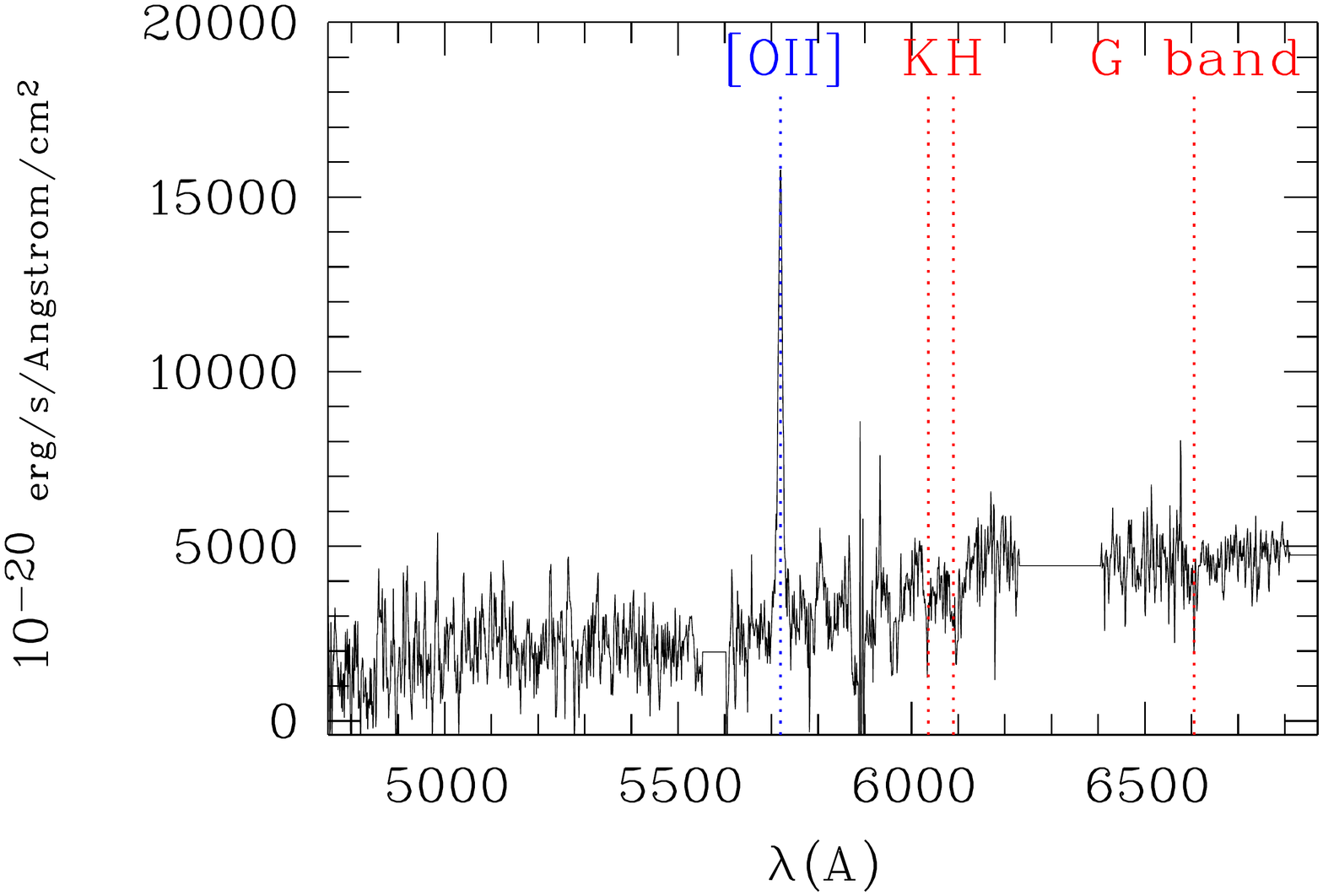}
    \includegraphics[angle=270,width=3.55in,clip=true]{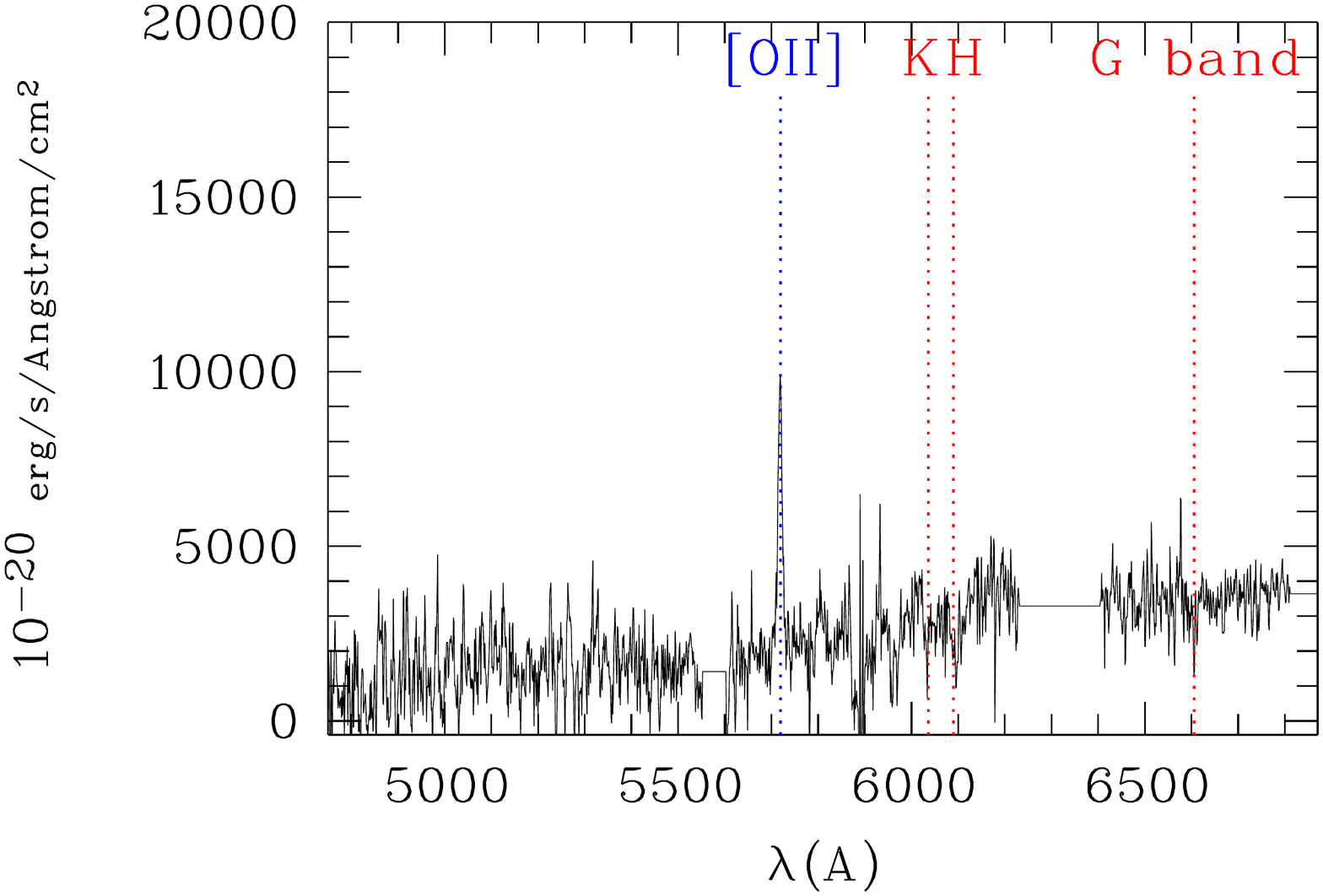}
    \includegraphics[angle=270,width=3.55in,clip=true]{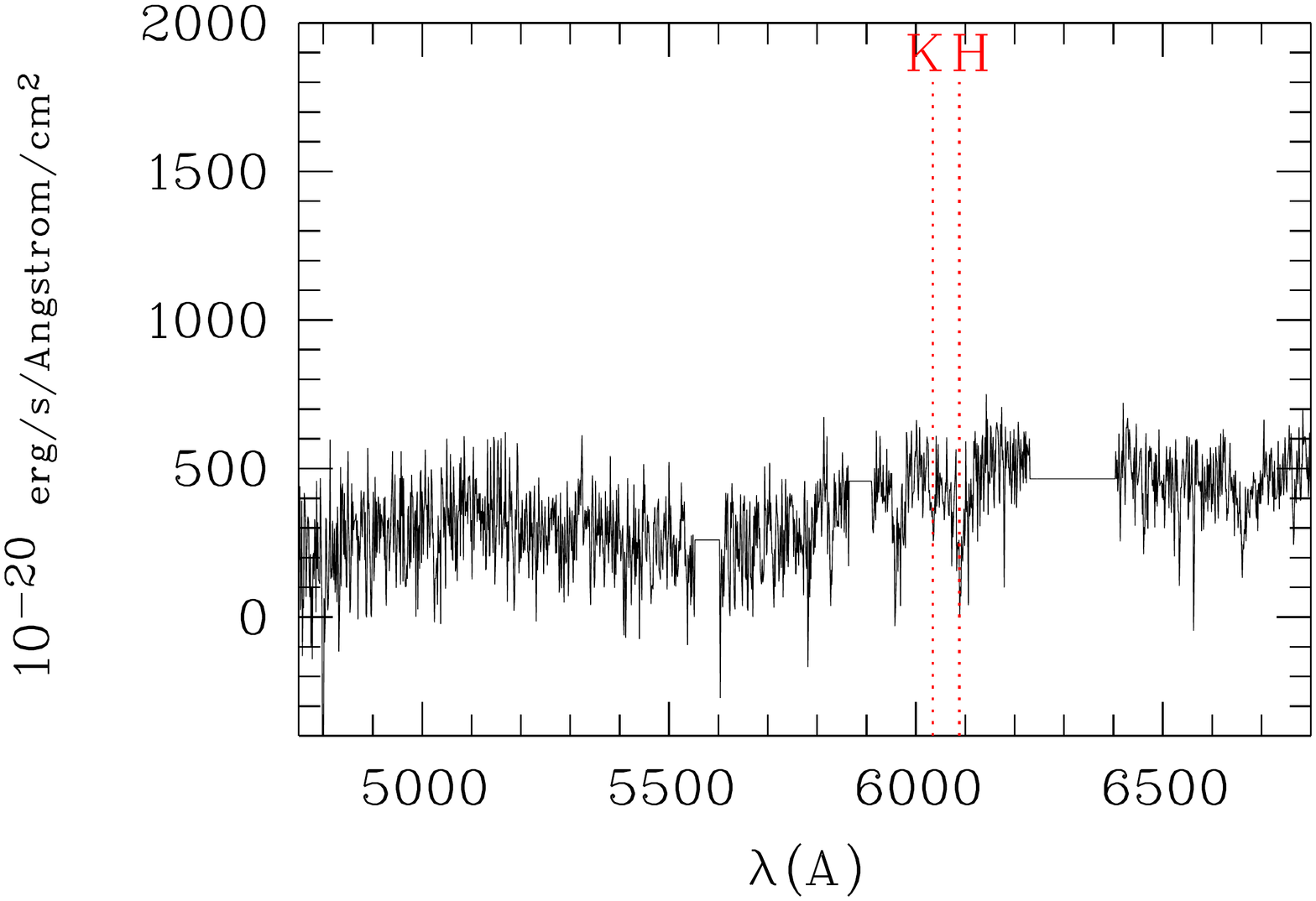}
    \caption{Top to bottom: (1) Region 1 diffuse light spectrum; (2) Region 1 diffuse light spectrum with the six non-CFHTLS emission 
line objects removed; (3) Region 2 diffuse light spectrum. }
  \label{fig:figDL2}
  \end{center}
\end{figure}

\subsection{Spectroscopy of six other non-CFHTLS sources}

We scanned the MUSE data searching for localised regions exhibiting emission lines without any counterparts in the CFHTLS object 
list. We detected such emission lines in six distinct arcsec-size regions (seeing dominated, clearly smaller than the ELDL source). 
They all show [OII] emission lines at approximately the cluster redshift (see Table 6) without being 
identified with a detected CFHTLS object. They also do not exhibit strong H$\beta$ lines. These sources are 
much smaller than the ELDL source. Two of these sources have a redshift slightly larger than 0.54. The other four are more clustered 
along the line of sight (see fig.~\ref{fig:galDL}) and have a velocity dispersion of 233 km/s. These four sources 
are probably not all inside the halo of the BCG given their spatial distribution (see figs.~\ref{fig:figDL} and ~\ref{fig:figDL2}), while 
the two others (at z$\geq$0.54) are not linked with the BCG given their redshift.

\begin{table}[t!]
  \caption{J2000 coordinates and redshifts of the emission line 
sources not detected in the CFHTLS data.}
\begin{tabular}{rrrr}
\hline
\hline
Id &  alpha        &  delta  &  redshift   \\
\hline
ELDL &  32.6656 & -5.9447  & 0.5339   \\
1    &  32.6638 & -5.9456  & 0.5477  \\  
2    &  32.6643 & -5.9437  & 0.5351  \\  
3    &  32.6662 & -5.9421  & 0.5578  \\ 
4    &  32.6649 & -5.9432  & 0.5352  \\  
5    &  32.6650 & -5.9427  & 0.5342  \\  
6    &  32.6651 & -5.9445  & 0.5334  \\
\hline
\end{tabular}
\label{tab:sampleemline}
\end{table}

\subsection{Spectroscopy of CFHTLS-detected diffuse light}

Figs.~\ref{fig:figDL} and ~\ref{fig:figDL2} show  trichromic g', i', and K$_s$ images of the cluster centre. We note that the ELDL source (red 
ellipse) is not visible in this figure because it is only visible in narrowband  images as previously stated. It shows the six non-CFHTLS 
emission line objects, and the ellipses we defined in fig.~\ref{fig:images2} for the i' band. These ellipses were grouped into two regions 
labelled 1 and 2 (see figs.~\ref{fig:figDL} and ~\ref{fig:figDL2}).

We extracted spectra of the diffuse light detected earlier in CFHTLS i' band images. This diffuse light is very 
low surface brightness, and it is therefore very difficult to reach a decent signal-to-noise ratio
in the spectroscopy; however, we were  able to extract exploitable spectra for the two defined regions (see figs.~\ref{fig:figDL} and 
~\ref{fig:figDL2}). We first excluded the places 
occupied by CFHTLS detected compact objects (galaxies) inside their CFHTLS ellipse. We then subtracted a background spectrum (defined in a circular 
empty region). This gave us the two spectra shown in figs.~\ref{fig:figDL} and 
~\ref{fig:figDL2}.

The first spectrum (region 1 in fig.~\ref{fig:figDL})  shows an [OII] emission line on top of an early-type spectrum characterised by clearly 
visible H$\&$K lines. This source is therefore dominated by old stars. There is no visible H$\beta$. This probably means that
H$\alpha$ is also weak and that following Kennicutt et al. (1998) the SFR is very low. Another explanation could be that this region 
is very dusty, thereby masking the H$\beta$ emission.  In order to know if the [OII] emission is due to the six compact 
regions described previously, we removed the contribution of these regions. Following this removal, the [OII] emission is still visible. This shows that region 1
probably has diffuse [OII] emission in addition to the emission from the six compact regions. The redshift of region 1 is 0.5345, 
identical to the shift of the BCG.

The second source (region 2) has a pure early-type spectrum (also dominated by old stars) without any detectable emission lines. Its redshift is 0.5340, 
150 km/s ahead of region 1.

\subsection{Emission lines around other galaxies in the field}

We also investigated the possible presence of ionised gas around other bright elliptical galaxies in the field as this subject is of prime interest
for star formation in clusters (see e.g. Donahue et al. 2015).  In fig.~\ref{fig:emDL2} we show the significance contours of the diffuse light detected
in the summed MUSE [OII] (5699-5729 A), H$\beta$ (7432-7471 A), and [OIII] (7655-7696 A) images. Except for the BCG, there are nearly no 
detections around the 
other visible galaxies. The only possible exception  comes from the two elliptical galaxies at ($\alpha$; $\delta$) = (02 10 40.26; -05 56 33) and 
(02 10 40.33; -05 56 36) toward the BCG north-east where some 2/2.5$\sigma$ small sources may be present. 

\section{Conclusions}

In this article, we studied a cluster detected in the XXL Survey with an apparent very high level of diffuse light.

- WIRCAM and MUSE data first allowed us to put the cluster characteristics on a firmer ground. It is a z$\sim$0.534 cluster of galaxies with old 
stellar populations and consists of a main component with a velocity dispersion of the order of 600 km/s and  an infalling low-mass group with 
a velocity dispersion of 170 km/s.

- We performed a Wavelet analysis of CFHTLS and WIRCAM images and found diffuse light inside a 60x180 kpc region. Detection was possible 
at all wavelengths, from u* band to K$_s$ band. The amount of diffuse light is equivalent to two BCGs in the i' band. To the best of our 
knowledge, this is the first detected cluster with such a large amount of diffuse light. The general orientation of the DL is 
north-east--south-west.

- SED fitting based on CFHTLS and WIRCAM data shows an early-type spectrum for the diffuse light, in contrast to the BCG which shows an Sd
type.

- Using MUSE spectroscopy, we first detected a new 13$\times$6 kpc source (ELDL source) only emitting in [OII], [OIII], and weakly in H$\beta$, with no 
detectable
continuum in any of the CFHTLS and WIRCAM data. Its redshift coincides with the cluster redshift. This source is very star-poor, with a 
low excitation parameter of 0.01. Its gas probably 
has half a solar metallicity, which is statistically lower than some other metallicities observed in other cluster ICL (see e.g. De Maio et al. 2015).
 
- We also detect six other small-scale emission line sources that are not detected in the CFHTLS images, but  are at the cluster redshift. 
Spatially, some of these sources are not located within the BCG  halo and are therefore distributed in the cluster halo.

- The diffuse light detected in CFHTLS and WIRCAM images near the BCG (region 1) exhibit some emission lines in addition to H\&K lines, 
while the other region (region 2) contains only absorption lines. Emission lines in region 1 are not only due to the six small-scale emission line sources,
but also to a more diffuse emitting component. This diffuse light is also at the cluster redshift.

Our conclusions are the following.  First,  the emission lines detected in the diffuse light are probably not due to an intense
star formation process because the estimated SFR in the diffuse light is relatively low and typical of values in an elliptical galaxy. The emission
lines,  therefore, probably originate from gas similar to that found in objects of medium metallicity and low excitation. It is also 
similar to the Croxall et al. (2009) HII regions in tidal dwarf galaxies. 

Second, the ICL star population is not very old, of the order of 2.3$\times$10$^9$ yrs, compared to the BCG population of 9.3$\times$10$^9$ yrs. 
This is also statistically younger than studies such as Melnick et al. (2012). In the cluster they studied, only $\sim$25$\%$ of the ICL stars
they detected are younger than 2 Gyrs. Most of the ICL in their cluster is composed of $\sim$8-10 Gyr-old stars (see their Fig.6). This can  be 
explained, however,  as the cluster Melnick et al. (2012)  consider has three bright interacting galaxies with  old stellar populations. This 
probably 
introduces a large number of old stars in the ICL (see their Fig. 2), while   our case has a very different configuration with a single BCG. This 
BCG is also forming a large number of stars, even more than the star-forming galaxies in the Puech et al. (2010) sample. This may appear to 
contradict the old age derived for the BCG population. A possible explanation would be a relatively recent episode of star formation 
in the BCG. Given the actual SFR, this burst would need at least 2 Gyr to generate enough young stars to account for a 
significant amount (a quarter) of the BCG stellar mass. This value is close to the age of the ICL stellar population. Therefore, 
this burst probably did not occur more than $\sim$2 Gyr ago. 

Third, the ICL has a slightly lower stellar mass (10$^{10.7}$ M$_\odot$) compared to the BCG (10$^{10.9}$ M$_\odot$), while it is clearly
a magnitude brighter in g', r', i', z' bands. This means that gas in the ICL dominates the stellar contribution in the optical (as is also the case 
in the ELDL source). This is at odds with other ICL spectral characterisations as in Melnick et al. (2012) where the young 
metal-poor component in ICL is less
than 1$\%$. This shows that we really are dealing with a cluster where matter ejection from galaxies is peculiar. Processes 
that may play a role in this matter ejection are ram pressure stripping, turbulent viscous stripping, or supernovae
winds. Other processes, which act principally on stars, are not dominant and not favoured (e.g. slow galaxy-galaxy interactions or mergers). 
Moreover, this ICL is not forming a large number of stars, at most on the order of 1 or 2M$_\odot$/yr in the ELDL source.

Fourth, we also know
that XLSSC 116 is not currently undergoing a major merger and only shows evidence of a minor and very low temperature substructure clump, 
likely falling onto the cluster from the rear side. Any other infalling galaxy structures must therefore already be in an advanced merging stage, and so do not
show as dynamically independent structures any more. In that case, any additional structures must have {passed} the second pericentre 
approach to be undetectable (as shown in Poole et al. (2007)). The time elapsed between an infalling group's
first virial crossing and the second pericentric approach obviously depends on the cluster-infalling structure mass ratio and on the impact 
parameter. However, assuming a value of 2 Gyr (the epoch of the potential burst of star formation in the BCG) for this timescale, Fig. 1 
of Poole et al. (2007) seems to require an intermediate impact parameter of the order of 0.15 or lower and, more importantly, a mass ratio close to 1 between the cluster
and the potentially infalling structure. Such a merger could have induced the observed shift between the XLSSC 116 BCG and the X-ray centre (e.g. Adami $\&$ Ulmer 2000). 
This potentially implies a rapid motion of bright galaxies, including the BCG, inside the cluster
gas that induces a ram pressure stripping process. A mass ratio close to 1 is also ideal for 
maximising the energy and matter exchange process between the impacted galaxy structure and the impactor, thereby generating a large amount of 
detected ICL. Spectacular examples of this kind of interaction are presented in Fumagalli et al. (2014), but are 
generally smaller in size. 
In conclusion, this cluster was likely built by several overlapping mergers: an older merger of two comparable mass systems and another ongoing minor
merger. 

This is the first time that we have been able to spectroscopically constrain diffuse light in a distant cluster and this illustrates the potential
of MUSE observations for such studies. We need more observations of a larger sample of clusters in order to draw a statistical picture
of the diffuse light in clusters of galaxies.

\begin{acknowledgements}

XXL is an international project based around an XMM Very Large Programme surveying two 25 deg2 extragalactic 
fields at a depth of $\sim$5 $\times$ 10$^{-15}$ erg/cm$^2$/s in the [0.5-2] keV band for point-like sources. The XXL 
website is http://irfu.cea.fr/xxl. Multiband information and spectroscopic follow-up of the X-ray sources are 
obtained through a number of survey programmes, summarised at http://xxlmultiwave.pbworks.com/.

Based on observations obtained with    
    MegaPrime/MegaCam, a joint project of CFHT and CEA/IRFU, at the
    Canada-France-Hawaii Telescope (CFHT) which is operated by the
    National Research Council (NRC) of Canada, the Institut National
    des Sciences de l'Univers of the Centre National de la Recherche
    Scientifique (CNRS) of France, and the University of Hawaii. This
    work is based in part on data products produced at TERAPIX
    available at the Canadian Astronomy Data Centre as part of the
    Canada-France-Hawaii Telescope Legacy Survey, a collaborative
    project of NRC and CNRS. This research has made use of the VizieR catalogue
    access tool, CDS, Strasbourg, France. This research has also made
    use of the NASA/IPAC Extragalactic Database (NED) which is
    operated by the Jet Propulsion Laboratory, California Institute of
    Technology, under contract with the National Aeronautics and Space
    Administration. Based on observations obtained with XMM-Newton, an ESA 
science mission with instruments and contributions directly funded by
ESA Member States and NASA. Based on observations made with the WHT telescope, operated on 
the island of La Palma by the Isaac Newton Group in the Spanish Observatorio del Roque de 
los Muchachos of the Instituto de Astrofísica de Canarias. 

We also gratefully acknowledge financial support from the Centre National d'Etudes Spatiales 
throughout many years of research. We also acknowledge financial support from ``Programme National de 
Cosmologie et Galaxies'' (PNCG) of CNRS/INSU, France and from Franco-Italian PICS. We thank 
V. Mainieri and L. Coccato for their help in reducing the MUSE data. We thank C. da Rocha for the
use of his wavelet code. We thank E. Slezak for his help during the first phases of this work. We
thank L. Chiappetti for a careful reading of the manuscript. We thank O. Ilbert for help in using the LePhare
code. We finally thank the referee for useful and detailed comments.

\end{acknowledgements}

\end{document}